\documentclass[11pt]{article}

\PassOptionsToPackage{hyphens}{url}
\usepackage[final]{acl}
\PassOptionsToPackage{hyphens}{url}
\makeatletter
\g@addto@macro{\UrlBreaks}{\UrlOrds}
\makeatother
\AtBeginDocument{%
  \def\UrlBreaks{\do\.\do\@\do\\\do\/\do\!\do\_\do\|\do\;\do\>\do\]%
    \do\)\do\,\do\?\do\&\do\=\do\#\do\~\do\-\do\+%
    \do\a\do\b\do\c\do\d\do\e\do\f\do\g\do\h\do\i\do\j\do\k\do\l%
    \do\m\do\n\do\o\do\p\do\q\do\r\do\s\do\t\do\u\do\v\do\w\do\x\do\y\do\z%
    \do\A\do\B\do\C\do\D\do\E\do\F\do\G\do\H\do\I\do\J\do\K\do\L%
    \do\M\do\N\do\O\do\P\do\Q\do\R\do\S\do\T\do\U\do\V\do\W\do\X\do\Y\do\Z%
    \do0\do1\do2\do3\do4\do5\do6\do7\do8\do9}%
}

\usepackage{times}
\usepackage{latexsym}
\usepackage{amsmath}
\usepackage{amssymb}

\usepackage[T1]{fontenc}
\usepackage[utf8]{inputenc}

\usepackage{microtype}

\usepackage{graphicx}
\usepackage{subcaption}
\usepackage{listings}
\usepackage{booktabs}
\usepackage{multirow}
\usepackage{pifont}
\usepackage{xstring}
\graphicspath{{../figures/}}

\usepackage{amsmath,amsfonts,bm}
\usepackage{booktabs} % For formal tables
\usepackage{url}
\usepackage{graphicx}
\usepackage{array}
\usepackage{color}
\usepackage{makecell}
\usepackage{multirow}
\usepackage{xspace}
\usepackage{colortbl}
\usepackage{subcaption}
\usepackage{algpseudocode}
\usepackage{setspace}
\usepackage{varwidth}
\usepackage{enumitem}
\usepackage{mdframed}
\usepackage{centernot}
\usepackage{textcomp}
\usepackage{tcolorbox}
\tcbuselibrary{breakable}
\usepackage{soul}
\usepackage{pifont}
\usepackage{graphics}
\usepackage{arydshln}

\usepackage{wrapfig}

\usepackage{hyperref}
\usepackage{url}

\usepackage{cuted,tcolorbox,lipsum}

\usepackage{listings}
\usepackage{xcolor}
\usepackage{twemojis}

\usepackage{caption}   % for \captionof
\newcommand{\start}[1]{\vspace{.3mm}\noindent{{\bf #1}.}}

\definecolor{amber}{rgb}{1.0, 0.75, 0.0}
\definecolor{applegreen}{rgb}{0.55, 0.71, 0.0}
\definecolor{treegreen}{rgb}{0.13, 0.54, 0.23}
\definecolor{LightCyan}{rgb}{0.88,1,1}
\definecolor{LightBlue}{RGB}{171, 227, 235}

\definecolor{green2}{HTML}{BFD8B6}
\definecolor{green3}{HTML}{E7F0E5}
\definecolor{greenarrow}{HTML}{1DB100}
\definecolor{red3}{HTML}{C82506}

\definecolor{gred}{RGB}{255,102,102}
\definecolor{gblue}{RGB}{51,102,255}
\definecolor{gyellow}{RGB}{244,180,0}
\definecolor{ggreen}{RGB}{15,157,88}
\definecolor{ggrey}{RGB}{115,115,115}
\definecolor{na}{gray}{0.9}

\definecolor{textRed}{RGB}{157,0,23}
\definecolor{textYellow}{RGB}{166,119,54}
\definecolor{textGreen}{RGB}{58,110,38}
\definecolor{textBlue}{RGB}{39,71,156}

\definecolor{LightYellow}{RGB}{255,250,208}
\definecolor{LightGreen}{RGB}{194,255,192}
\definecolor{LightBlue}{RGB}{187,236,251}
\definecolor{LightPurple}{RGB}{224,223,255}
\definecolor{LightGrey}{RGB}{225,225,225}

\definecolor{OrangeRed}{rgb}{1.0, 0.27, 0.0}
\definecolor{Orange}{rgb}{1.0, 0.5, 0.3}
\definecolor{midnightgreen}{rgb}{0.0, 0.29, 0.33}
\definecolor{darkgreen}{rgb}{0.0, 0.42, 0.24}
\definecolor{purple}{RGB}{134, 45, 250}

\definecolor{diagramRed}{RGB}{246,193,193}
\definecolor{diagramPurple}{RGB}{224,224,253}
\definecolor{diagramOrange}{RGB}{244,222,176}

\definecolor{darkgreen}{HTML}{006400}

\newcommand{\cwm}{\texttt{CWM-32B}}
\newcommand{\qwenEightB}{\texttt{Qwen3-8B}}
\newcommand{\qwenThirtyTwoB}{\texttt{Qwen3-32B}}
\newcommand{\qwenNextEightyB}{\texttt{Qwen3-Next-80B-A3B-Instruct}}
\newcommand{\qwenNext}{\texttt{Qwen3-Next-80B-A3B}}
\newcommand{\opus}{\texttt{Claude-Opus-4.6}}

\title{Steer, Don’t Solve: \\ Training Small Critic Models for Large Coding Agents}

\author{
  \textbf{Shubham Gandhi*}\thanks{Equal contribution.},
  \textbf{Yiqing Xie*},
  \textbf{Atharva Naik},
  \textbf{Ruichen Zhu},
  \textbf{Carolyn Rosé}
\\
  Carnegie Mellon University
\\
  \small{\texttt{\{srgandhi, yiqingxi, arnaik, rzhu3, cp3a\}@andrew.cmu.edu}}
}

\begin{document}
\maketitle
\begin{abstract}
Coding tasks are typically complicated and require multiple capabilities, ranging from high-level planning to low-level implementation. While coding agents are optimized for the joint capabilities, individual capabilities such as high-level planning may have different optima and remain a major bottleneck.
To address this challenge, we train a separate critic model that is \emph{specialized in high-level planning} to steer the coding agent in inference.
We construct SFT and DPO data to train the critic model to identify errors made by the coding agent and provide correct and clear high-level guidance without generating concrete actions.
Experiments show that our fine-tuned 4B and 8B critic models significantly improve the performance of 6 larger coding agents (e.g., improving the resolved rates of \texttt{GLM-4.7-Flash-30B-A3B} and \texttt{GPT-OSS-120B} by 16.0\% and 14.4\% on SWE-Bench Verified).
The critic model also reduces the total inference costs for some coding agents by solving tasks in fewer steps (e.g., reducing the per-example inference cost for \texttt{GPT-OSS-20B} from \$0.07 to \$0.03).
\footnote{Code: \url{https://github.com/shubhamrgandhi/critic-training}.}
% End-to-end code agent training is resource-intensive and plateaus on the strategy-level reasoning needed to resolve code issues, since jointly optimizing code-level execution and strategy-level reasoning leaves the latter underdeveloped. Instead, we freeze the agent and add a \emph{critic model} to supply that signal. Prior code critics are \emph{post-hoc}, scoring completed trajectories rather than steering the agent; we instead train a small critic that provides \emph{intra-trajectory} feedback via Supervised Fine-Tuning. On SWE-Bench Verified, a critic trained on \cwm{} trajectories transfers to two unseen agents (gains of $+3.0$ to $+3.8$ points), and adding target-agent trajectories to the corpus increases the gain to $+3.8$ on \cwm{} and $+4.4$ to $+5.2$ on two Qwen agents, at $30$-$92\times$ lower critic cost than a strong teacher. On \qwenNext, the critic-guided system is both more accurate ($25.2\%$ vs.\ $20.8\%$) and cheaper ($\$0.04$ vs.\ $\$0.11$) than the agent alone, because the critic also shortens trajectories. Our results show that a small, well-trained critic is a practical complement to scaling agent training.
% Data and models: \url{https://huggingface.co/collections/shubhamrgandhi/critic-training-for-code-agents}.
\end{abstract}

\begin{figure}[t]
    \centering
    \includegraphics[width=\linewidth]{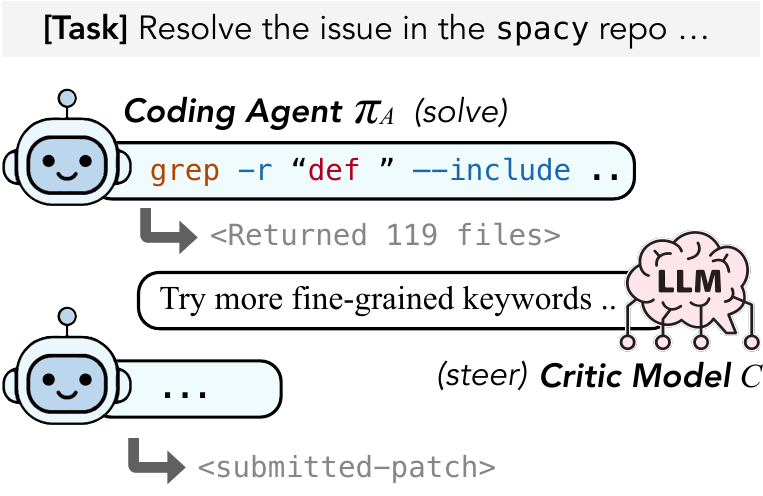}
    \caption{Inference setup. A trained critic model $C$ provides guidance to a frozen coding agent $\pi_A$ for every $k$ steps. The agent is responsible for all concrete actions and the critic model inspects the current trajectory and generates high-level feedback for its next steps.}
    \label{fig:inference}
\end{figure}

\section{Introduction}
\label{sec:introduction}

% \yiqing{We should probably mention the performance saturation issue here (by citing R2EGym's plot). Then we briefly talk about the code agent is optimized for multiple abilities and when the performance saturated, the single abilities are still far from perfect, which motivates training a critic model specialized for planning.}
% \shubham{Rewrote around the saturation argument with three planning-gap citations.}
% \yiqing{Be more concise about the optimum doesn't align problem}
Coding agents have been widely used in real-world programming tasks~\citep{ICLR2024_edac78c3,swe-agent}, and post-training has proven effective for coding agent training~\cite{pmlr-v267-pan25g,yang2025swesmith,xie2026hybridgymtrainingcodingagents}, where the model is optimized for the combination of multiple abilities, including high-level reasoning (e.g., troubleshooting, search, planning) and low-level implementation (executing tool calls correctly).

% However, empirically, this training is resource-intensive and the performance plateaus quickly for base models with limited size. For instance, the issue-solving performance of a 14B model stops growing after training on only roughly 800 trajectories~\cite{jain2025regym}.
Theoretically, the optima for the joint capabilities could be different from the optima for each individual capability, such as high-level planning.
Empirically, during post-training, multiple studies have shown that the planning ability of a coding agent is still far from perfect after reaching the plateau on the training data~\cite{xie2026hybridgymtrainingcodingagents,wang2026rubric}.
For example, about 65\% of coding agent failures come from flawed reasoning, not wrong actions \citep{liu2025empiricalfailures}.
A possible explanation is that integrating both high-level and low-level knowledge in training makes the learned patterns overly complex, which undercuts the coding agent's ability to learn generalizable patterns and limits the overall performance. 

To address this challenge, we shift part of the high-level planning role to a separate, specialized \textbf{critic model}, which steers the main coding agent by providing feedback during inference.
The idea is inspired by research on critic and adviser models in other NLP domains such as question-answering~\cite{gou2023critic}, factuality~\cite{xie-etal-2025-factuality}, and mathematical problems~\cite{ICLR2024_aca97732}.
Compared to training the coding agent itself, we highlight the advantages of training a critic model as follows:
First, the critic model improves the overall performance by complementing the planning ability of the coding agent.
Second, critic model training is cost-efficient in terms of both data curation and computational resources.
\S\ref{sec:method} shows that we can re-use existing coding-agent environments for critic model training, without building new training tasks or environments.
\S\ref{sec:main_results} shows that to improve the overall performance, we can train a small critic model instead of training the large coding agent, which requires fewer computational resources.
Finally, the use of critic models is highly adaptive and flexible.
We can provide critiques to any frozen coding agent at any step with any frequency.

In this paper, we design a two-stage training framework to train the critic models.
The first training stage (\autoref{fig:training}) uses supervised fine-tuning (SFT), where we distill from a strong teacher model (e.g., \opus{}).
We explicitly require the target critiques to detect the major errors made by the coding agent and provide corresponding high-level guidance to the next steps, without containing any concrete implementation such as specific tool-call commands.
As shown in our analyses, there is still a substantial gap between the overall guidance generated by the SFT student model and the teacher critic model.
Therefore, we present the second training stage with direct preference optimization (DPO), where we construct preference pairs based on the quality of the guidance in the critiques.
Specifically, we design two metrics: correctness and clarity.
Correctness requires the critiques to ground to the previous error categorization results and suggest the correct direction for future steps.
Clarity checks whether the critiques provide clear instructions without ambiguity or under-specification.

Experiments on SWE-Bench Verified~\cite{ICLR2024_edac78c3} show that our trained critic model improves the resolved rate of six different coding agent models by $5.6\%$--$16.0\%$, without re-training the coding agents. 
We further demonstrate that our training can transfer across different programming languages and coding agents in training and evaluation.
For instance, we improve the performance on Multi-SWE-Bench~\cite{zan2025multiswebench} by $2.6\%$--$3.6\%$ on 7 unseen languages.
We also conduct detailed analyses to demonstrate the effectiveness of the SFT stage, DPO stage, and the design choice of training on high-level critiques only.
Finally, we provide qualitative analyses and case studies on the critiques generated by our models.

% \start{Contributions}
% (1) to our knowledge, the first study of training critic models for coding agents, as a separation-of-concerns alternative to scaling the agent; (2) an analysis of which design choices about feedback form (detailed vs.\ high-level prompts) transfer from teacher to student; and (3) empirical validation on SWE-Bench Verified showing that a critic trained on a single agent's trajectories transfers to unseen agent models (gains of $2.6$--$3.6$ points), while our best mixed-corpus configurations improve all six agent models by $5.6$--$16.0$ points.

\section{Methodology}
\label{sec:method}

% \yiqing{Motivate critic models for coding agents. Probably make it shorter.}
Long-horizon coding agents drift: they over-commit to early hypotheses, repeat failed commands, or allocate steps to subgoals in a suboptimal manner~\cite{gandhi2025sweprm,wang2026rubric}. To overcome this challenge, we leverage a critic as a second model to inspect the trajectory at runtime in order to alert the model to such drift.
Empirically, pairing a frozen coding agent with a strong proprietary model such as \texttt{Claude-Opus-4.6}~\cite{claude46} as a critic substantially improves the performance but also requires a higher total cost.
For instance, it improves the resolved rate of \texttt{Qwen3-32B} on SWE-Bench-Verified from 8.8\% to 26.8\%, but also increases the per-instance cost from \$0.05 to \$0.28.

To achieve a better balance between cost and performance, in this section, we train a smaller open-source critic model to guide the coding agent during inference.
Specifically, we first introduce the formal definitions (\S\ref{sec:critic_models}) and the design of the critique (\S\ref{sec:critique_design}). Then we design training data for the SFT (\S\ref{sec:sft}) and DPO stages (\S\ref{sec:dpo}).

\subsection{Formal Definitions}
\label{sec:critic_models}
\start{Coding Agent}
Given a coding task and a docker environment in which to perform the task, the coding agent has access to a set of tools (e.g., the bash command-line tool) that can be called through tool-call actions.
In problem-solving, the agent generates a trajectory as the solution 
$\tau_{1:t} = (a_1, o_1, \ldots, a_t, o_t)$,
where $a_i$ is the agent's action and $o_i$ the corresponding environment observation.
We limit the number of steps with a fixed budget $t \leq B$.
In our setting, $\pi_A$ is held frozen throughout and we only train the critic model $C_\theta$. 
% Scaffold mechanics and task-instance specification are detailed in App.~\ref{app:method_prelim}.
% The coding agent $\pi_A$ in Figs.~\ref{fig:inference} and~\ref{fig:training} is the Mini-SWE-Agent scaffold~\cite{swe-agent}\footnote{\url{https://github.com/SWE-agent/mini-swe-agent}}, which has access to the bash command-line tool.

\start{Critic Models for Coding Agents}
Formally, we pair the coding agent $\pi_A$ with a critic model $C$ at inference time.
For every $k$ agent steps (we use $k=\{5,10\}$ in our experiments), the critic model inspects the task and the current trajectory prefix $\tau_{1:t}$. 
To preserve continuity across interventions, we also include the feedback the critic produced at earlier intervention points. Let $h_t = (f_k, f_{2k}, \ldots, f_{t-k})$ denote the critic's prior feedback history before step $t$. We form the critic input $z_t = \mathrm{Serialize}(\tau_{1:t}, h_t, x)$, which fits the trajectory to the context window limit $x$ by preserving the start of the trajectory (system prompt and task context) and the most recent steps, including the final user message, and finally truncating the intermediate steps that exceed the context limit.
We sample feedback $f_t \sim C(\cdot \mid z_t)$, which is appended to the agent's context for the next step.
% We evaluate intervention intervals $k \in \{5,10\}$.

\subsection{Designing the Critique Structure}
\label{sec:critique_design}
We design the structure of the critique as follows: it first outputs the major errors made by the coding agent and the recovery action, where the error categories are pre-defined.
Then it outputs the overall guidance, which contains only high-level feedback instead of the concrete tool call actions.
Finally, it outputs a reminder when the agent is close to using up the step limit.
We provide the prompt for critique generation in App. \ref{app:high_level_prompt}.

% \start{What should a critic say}
% The ``High-level only'' constraint in Fig.~\ref{fig:inference} is enforced through the critique design. Following the observation in §\ref{sec:critic_models} that the trained critic should not inherit the teacher's action-level recipes, the critique we give to the agent has three components: (i) it organizes feedback over the 12-category trajectory-level error taxonomy of \citet{gandhi2025sweprm} (e.g., \emph{Step Repetition}, \emph{Hallucinations}, \emph{Task Derailment}); (ii) it gives \emph{high-level feedback} rather than prescribing explicit actions; and (iii) it adapts as the agent's budget runs out. (i) is inherited from the upstream prompt; (ii) and (iii) are the design choices we study, and they are enforced at different stages: (ii) is a \emph{supervision} property baked in at training time by collecting teacher feedback under the high-level prompt, so the trained critic inherits it; (iii) is an \emph{inference-time} mechanism, budget-aware instructions appended to the trained critic's own input late in the trajectory and never used during data collection. The three paragraphs below introduce each component and the mechanism behind it.

\start{Error Detection for the Coding Agent}
Multiple existing studies show that coding agents' capability to recover from errors is still limited~\cite{wang2026rubric,gandhi2025sweprm}.
One example is the ``stuck-in-loop'' behavior, where the agent keeps generating the same tool call action even if it fails every time. 
To improve error recovery, we prompt the critic model to detect major errors made by the coding agent.
Inspired by SWE-PRM~\cite{gandhi2025sweprm}, we pre-define 12 trajectory-level error categories covering \emph{Specification}, \emph{Reasoning}, and \emph{Coordination} errors. The complete list is provided in App.~\ref{app:critic_prompts}.
For each category, the critic model reports whether the error is detected, the supporting evidence, recovery strategies, and a task-status verdict showing whether the error is not significant and the agent is still on track, or the error is serious and needs correction or critical intervention.

\start{High-level Feedback Only, NOT Concrete Actions}
The motivation for critic model training is to separate the high-level planning ability to a \textbf{specialized} model.
As a result, we explicitly prompt the critic model NOT to suggest specific technical implementation or code-level changes, such as outputting the exact tool call command.
To improve correctness, we also require the suggested recovery strategies or overall guidance to be grounded in previous error detection results.
We additionally prompt the critic model to keep the guidance high-level (1-2 sentences per detected error), without giving too complicated instructions to coding agents that cannot be completed in a few steps. 
The prompt thereby restricts the teacher to high-level signals such as whether the agent is making progress, looping, relying on an unsupported assumption, or pursuing the wrong subgoal (App.~\ref{app:critic_prompts}).

\start{Budget Awareness}
A high-level critic that ignores the budget will keep recommending exploration even when the agent is close to running out of steps, leaving useful in-progress work unsubmitted. Therefore, in inference, once the agent passes step $T$ (e.g., we use $T{=}100$ of the $B{=}150$-step budget), we append short budget-aware instructions that state the steps remaining and recommend the coding agent to make a submission as soon as possible (App.~\ref{app:step_aware}).

\begin{figure}[!t]
\centering
\includegraphics[width=\linewidth]{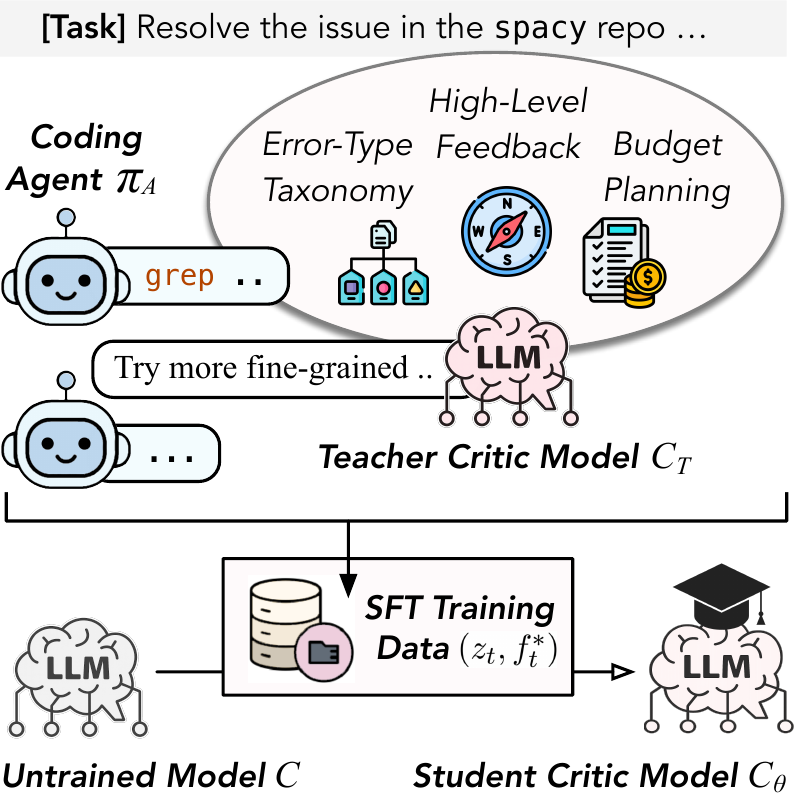}
\caption{SFT training pipeline for critic models. Based on our analysis (\S\ref{sec:critique_design}), the critic contains error detection, budget planning, and only provides high-level feedback instead of concrete actions. We run the frozen coding agent $\pi_A$ on disjoint training tasks and query the teacher critic $C_T$ every $k$ steps. We collect (trajectory prefix, feedback) pairs $(z_t, f_t^*)$ to train a critic $C_\theta$ by supervised fine-tuning (SFT).}
\label{fig:training}
\end{figure}

\subsection{Distilling Critic Models with SFT}
\label{sec:sft}
\start{SFT Data Construction}
As show in \autoref{fig:training}, we prompt the teacher critic model $C_T$ to provide guidance to a frozen coding agent $\pi_A$ every $k$ steps based on the current trajectory $z_t$ on the training dataset $\mathcal{TR_{SFT}}$ (as described in §\ref{sec:setup}) and fine-tune a student critic model $C_\theta$ with the teacher's outputs as targets.

Based on the prompt design introduced in \S\ref{sec:critique_design}, we prompt the teacher model to first detect errors made by the coding agent and then provide overall guidance, with budget-aware alerts. We additionally require it to only generate high-level feedback rather than action-level prescriptions.
We filter out malformed teacher responses (empty completions or responses that drop out of the critique format) before training (\S\ref{sec:setup}).
Note that we do NOT filter training critiques based on whether the task is successfully completed, which depends more on the coding agent itself rather than the critic model.

% the supervision source is the \emph{high-level prompt} of §\ref{sec:critique_design}, which constrains the teacher critic $C_T$ to high-level feedback rather than action-level prescriptions; sampling $f_t^* \sim C_T(\cdot \mid z_t)$ under this prompt is what produces our training labels. Second, we collect feedback along trajectories that the agent has actually produced under critic guidance, rather than from off-policy rollouts: we run the agent $\pi_A$ end-to-end with $C_T$ intervening every $k$ steps, so each prefix $\tau_{1:t}$ already reflects the agent's reaction to prior critic interventions $h_t$. This matches the distribution the trained critic will face at inference, where the prefix will be shaped by its own earlier feedback.
% We use training tasks that are disjoint at the instance level from the evaluation benchmark (R2E-Gym tasks in SWE-Bench repositories; described in §\ref{sec:experimental_setup}), so the critic is never exposed to test-time issues during training. As an ablation, we also collect a parallel corpus under the detailed prompt.

\start{Critic Model Training with SFT}
As shown in \autoref{fig:training}, we distill $C_\theta$ from $C_T$ by supervised fine-tuning (SFT) with a standard conditional language modeling objective, maximizing likelihood:
\begin{equation}
    \max _{\theta} \mathbb{E}_{(z_t, f_t^*) \sim \mathcal{TR_{SFT}}} \log \mathbb{P}_{\theta}(f_t^* \mid z_t).
\label{eq:sft_loss}
\end{equation}

The trained critic model $C_\theta$ is then paired with $\pi_A$ at inference..

\subsection{Improving Critic Models with DPO}
\label{sec:dpo}
\start{The Remaining Gap after Distillation}
We empirically study the gap between the student and teacher critic models after SFT (\S\ref{sec:sft}).
As shown in App. \ref{app:analyses}, the major gap exists in the \textbf{overall guidance} provided by the critic, instead of the error categorization.
Among the 132 critiques we sampled, 67\% of the teacher model's critiques both point to the correct direction and contain specific guidance for the next steps.
In contrast, only 29\% of the SFT student model's critiques are both correct and specific. 53\% point to the correct direction but are under-specified. 18\% suggest the wrong steps to take.
We use an LLM judge for the judgment. We choose a different model from the teacher model to avoid self-bias~\cite{llmjudge}.
We thus train the critic model for the second stage, focusing on generating high-quality overall guidance.

\start{DPO Data Construction}
We design two metrics to judge the quality of the guidance: (1) correctness, where the guidance fixes the serious error made by the coding agent (if any) and assists problem-solving; and (2) clarity, where the guidance is not ambiguous or under-specified.
We adopt direct preference optimization (DPO) to improve these two metrics.

\begin{figure}[!t]
\centering
\includegraphics[width=\linewidth]{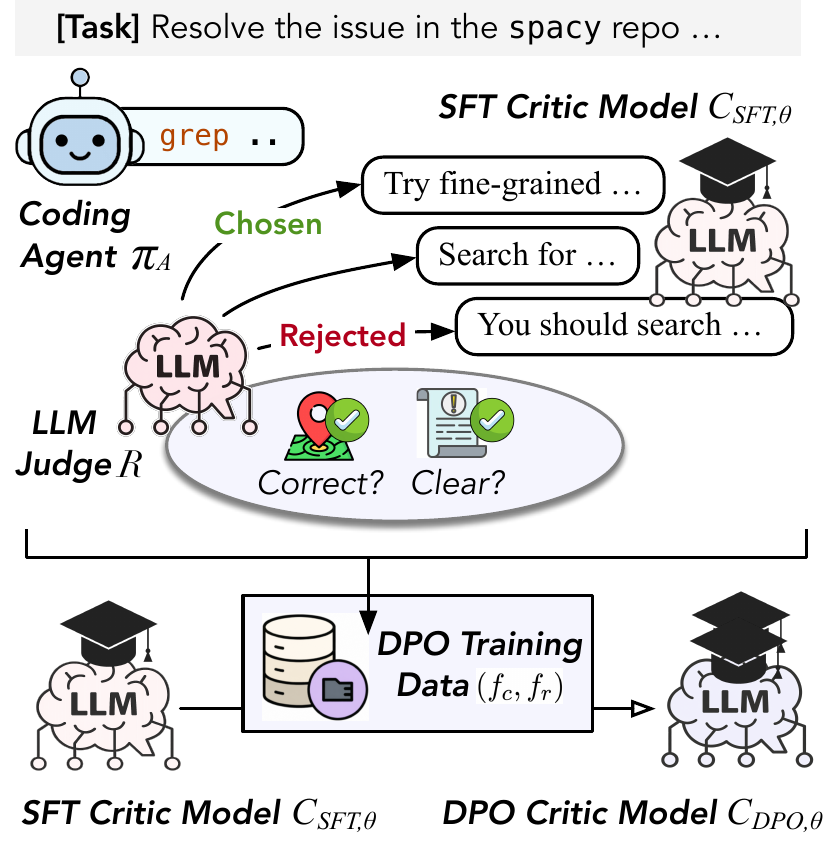}
\caption{DPO training pipeline for critic models. 
Given a critic model checkpoint $C_{SFT,\theta}$, for every $k$ steps produced by the coding agent $\pi_A$, we use $C_{SFT,\theta}$ to generate $N=10$ critiques, and prompt an LLM judge model $R$ to select the best and worse critique, which forms (chosen, rejected) pairs $(f_c, f_r)$ to train $C_{SFT,\theta}$ by direct preference optimization (DPO).
We only keep a $(f_c, f_r)$ pair if the chosen critique is both correct and provides clear instruction and the rejected critique misses any of the criteria.
}
\label{fig:training_dpo}
\end{figure}

As shown in \autoref{fig:training_dpo}, we apply the coding agent $\pi_A$ on the training dataset $TR_{DPO}$. For every $k$ steps, we use the SFT critic model $C_{SFT,\theta}$ to generate $N=10$ critiques, with the previous trajectory $z_t$ as context. Then we prompt an LLM judge to select the best and worst ones based on the correctness and clarity of the overall guidance.
This forms one (chosen, rejected) training pair $(f_c, f_r)$ for every set of critiques. We only keep the training pairs where the error categorization does not have major error, the chosen critique is both correct and clear, and the rejected critique is either not correct or not clear.

\start{Critic Model Training with DPO}
As shown in \autoref{fig:training_dpo}, we initialize from the SFT critic model $C_{SFT,\theta}$ and conduct DPO training with the following classification loss:
\begin{equation}
\begin{aligned}
    \max _{\mathcal{\theta}} 
    \mathbb{E}_{\left(z_t, f_c, f_r\right) \sim \mathcal{TR}_{DPO}}
    & \left[\right.
    \log \sigma
    \left(\right. \beta 
    \log \frac{\pi_\mathcal{\theta}\left(f_c \mid z_t\right)}{\pi_{\text {ref}}\left(f_c \mid z_t\right)} \\
    & -\beta 
    \log \frac{\pi_\mathcal{\theta}\left(f_r \mid z_t\right)}{\pi_{\text {ref}}\left(f_r \mid z_t\right)}
    \left.\right)
    \left.\right],
\end{aligned}
\end{equation}

where $\sigma$ is the Sigmoid function. The reference policy $\pi_{\text {ref}}$ is computed by the SFT checkpoint $C_{SFT,\theta}$ and is frozen throughout training.

\section{Experiments}
\label{sec:experiments}
This section evaluates the performance of our trained critic models on downstream coding tasks. 
\S\ref{sec:setup} and \S\ref{sec:details} introduces the experimental setup and implementation details.
\S\ref{sec:main_results} demonstrates the effectiveness of our critic models of different sizes when combined with six different coding agent models.
\S\ref{sec:detailed_analyses} presents detailed analyses, including the transferability of our critic models across programming languages and training corpus, the cost-performance trade-off of critic models, and the verification of our design choices that the critiques should only contain high-level guidance, not concrete actions.
\S\ref{sec:qualitative} provides qualitative analyses and case studies for the generated critiques.

\subsection{Experimental Setup}
\label{sec:setup}

\start{Benchmarks}
We primarily evaluate on SWE-Bench Verified~\citep{ICLR2024_edac78c3}, a benchmark of 500 human-verified GitHub issues drawn from 12 open-source Python repositories. 
We also conduct evaluation on its multilingual variant, Multi-SWE-Bench~\cite{zan2025multiswebench}, to demonstrate the transferability of our method across programming languages.
For each instance, the agent is given the issue's problem statement and access to the pre-issue repository snapshot, and must edit the codebase to resolve the issue. An instance is counted as resolved only if the resulting patch satisfies two hidden test sets: \texttt{FAIL\_TO\_PASS} tests that target the reported bug, and \texttt{PASS\_TO\_PASS} tests that guard against regressions in unrelated code.

% Main results table on SWE-Bench Verified across agent models.
% All arrows are vs no-critic.
% \dup = up & good (green); \ddn = down & bad (red).
% \dupR = up & bad (red); \ddnG = down & good (green); \dzero = zero (green).
% \approxzero = ~0.00 (display when actual >$0 but rounds to 0.00).

% \providecommand{\dup}[1]{\textcolor{darkgreen}{\scriptsize($\uparrow$#1)}}
% \providecommand{\ddn}[1]{\textcolor{red}{\scriptsize($\downarrow$#1)}}
% \providecommand{\dupR}[1]{\textcolor{red}{\scriptsize($\uparrow$#1)}}
% \providecommand{\ddnG}[1]{\textcolor{darkgreen}{\scriptsize($\downarrow$#1)}}
% \providecommand{\dzero}{\textcolor{darkgreen}{\scriptsize(0.0)}}

\providecommand{\dup}[1]{\textcolor{darkgreen}{\xspace($\uparrow$#1)}}
\providecommand{\ddn}[1]{\textcolor{red}{\xspace($\downarrow$#1)}}
\providecommand{\dupR}[1]{\textcolor{red}{\xspace($\uparrow$#1)}}
\providecommand{\ddnG}[1]{\textcolor{darkgreen}{\xspace($\downarrow$#1)}}
\providecommand{\dzero}{\textcolor{darkgreen}{\xspace(0.0)}}
\providecommand{\approxzero}{${\sim}0.00$}
% Cell-level shading (so the Coding Agent column stays unshaded).
\providecommand{\gc}{\cellcolor{applegreen!10}}
\providecommand{\yc}{\cellcolor{gray!20}}

\newcommand{\smallpad}{\hspace{0.9em}}
\newcommand{\largepad}{\hspace{3em}}

\begin{table*}[!t]
\centering
\resizebox{0.9\textwidth}{!}{%
\begin{tabular}{@{}l@{\hspace{1em}}!{\color{gray!45}\vrule}@{\hspace{1em}}lcccc@{}}
\toprule
\bf Coding Agent & \bf Critic Model & \bf Resolved \% & \bf Localized \% & \bf Non-Empty \% & \bf Non-Loop \% \\

\midrule
\multirow{7}{*}{\texttt{Qwen3-32B}}
  & No Critic          & 8.8 & 49.2 & 74.8 & 92.0 \\
  & \texttt{Qwen3-4B}  & 10.2\,\dup{1.4} & \textbf{55.4}\,\dup{6.2} & \textbf{90.0}\,\dup{15.2} & \textbf{95.2}\,\dup{3.2} \\
  & \gc + SFT          & \gc 11.4\,\dup{2.6} & \gc 52.4\,\dup{3.2} & \gc 74.6\,\ddn{0.2} & \gc 90.4\,\ddn{1.6} \\
  & \gc + SFT + DPO    & \gc \textbf{14.4}\,\dup{5.6} & \gc 52.2\,\dup{3.0} & \gc 69.2\,\ddn{5.6} & \gc 95.0\,\dup{3.0} \\
  & \texttt{Qwen3-8B}  & 10.0\,\dup{1.2} & 49.4\,\dup{0.2} & 73.0\,\ddn{1.8} & 93.0\,\dup{1.0} \\
  & \gc + SFT          & \gc 13.8\,\dup{5.0} & \gc 49.8\,\dup{0.6} & \gc 65.8\,\ddn{9.0} & \gc 93.0\,\dup{1.0} \\
  & \yc \texttt{Opus 4.6} (upper bound) & \yc \underline{26.8}\,\dup{18.0} & \yc \underline{67.8}\,\dup{18.6} & \yc 82.4\,\dup{7.6} & \yc 92.4\,\dup{0.4} \\

\midrule
\multirow{7}{*}{\shortstack[l]{\texttt{Qwen3-Next-}\\\texttt{80B-A3B}}}
  & No Critic          & 20.0 & 69.8 & 92.6 & 95.0 \\
  & \texttt{Qwen3-4B}  & 20.2\,\dup{0.2} & 70.4\,\dup{0.6} & \textbf{99.8}\,\dup{7.2} & 97.8\,\dup{2.8} \\
  & \gc + SFT          & \gc 24.2\,\dup{4.2} & \gc 76.8\,\dup{7.0} & \gc 99.2\,\dup{6.6} & \gc 98.2\,\dup{3.2} \\
  & \gc + SFT + DPO    & \gc \textbf{26.2}\,\dup{6.2} & \gc \textbf{78.4}\,\dup{8.6} & \gc 99.2\,\dup{6.6} & \gc \textbf{98.4}\,\dup{3.4} \\
  & \texttt{Qwen3-8B}  & 20.6\,\dup{0.6} & 69.4\,\ddn{0.4} & 97.6\,\dup{5.0} & 94.8\,\ddn{0.2} \\
  & \gc + SFT          & \gc 25.2\,\dup{5.2} & \gc 75.6\,\dup{5.8} & \gc \textbf{99.8}\,\dup{7.2} & \gc 96.2\,\dup{1.2} \\
  & \yc \texttt{Opus 4.6} (upper bound) & \yc \underline{38.2}\,\dup{18.2} & \yc \underline{81.4}\,\dup{11.6} & \yc \underline{100.0}\,\dup{7.4} & \yc \underline{99.4}\,\dup{4.4} \\

\midrule
\multirow{5}{*}{\texttt{GPT-OSS-20B}}
  & No Critic          & 3.0 & 11.8 & 58.4 & 1.4 \\
  & \texttt{Qwen3-4B}  & 6.8\,\dup{3.8} & 36.6\,\dup{24.8} & 98.8\,\dup{40.4} & 3.0\,\dup{1.6} \\
  & \gc + SFT          & \gc 9.8\,\dup{6.8} & \gc 52.6\,\dup{40.8} & \gc \textbf{99.6}\,\dup{41.2} & \gc 5.0\,\dup{3.6} \\
  & \gc + SFT + DPO    & \gc \textbf{14.8}\,\dup{11.8} & \gc \textbf{55.0}\,\dup{43.2} & \gc \textbf{99.6}\,\dup{41.2} & \gc \textbf{5.6}\,\dup{4.2} \\
  & \gc \texttt{Qwen3-8B} + SFT & \gc 13.0\,\dup{10.0} & \gc 52.0\,\dup{40.2} & \gc 99.4\,\dup{41.0} & \gc 5.4\,\dup{4.0} \\

\midrule
\multirow{5}{*}{\shortstack[l]{\texttt{GLM-4.7-}\\\texttt{Flash-30B-A3B}}}
  & No Critic          & 21.6 & 34.8 & 36.0 & 54.8 \\
  & \texttt{Qwen3-4B}  & 29.8\,\dup{8.2} & 62.2\,\dup{27.4} & 70.4\,\dup{34.4} & 67.2\,\dup{12.4} \\
  & \gc + SFT          & \gc 35.2\,\dup{13.6} & \gc 69.6\,\dup{34.8} & \gc 79.6\,\dup{43.6} & \gc 74.2\,\dup{19.4} \\
  & \gc + SFT + DPO    & \gc 35.8\,\dup{14.2} & \gc 71.0\,\dup{36.2} & \gc 79.2\,\dup{43.2} & \gc \textbf{75.0}\,\dup{20.2} \\
  & \gc \texttt{Qwen3-8B} + SFT & \gc \textbf{37.6}\,\dup{16.0} & \gc \textbf{73.4}\,\dup{38.6} & \gc \textbf{83.4}\,\dup{47.4} & \gc 70.6\,\dup{15.8} \\

% \midrule
% \multirow{2}{*}{\shortstack[l]{\texttt{GPT-OSS-120B}\\(low reasoning)}}
%   & No Critic          & 12.8 & 42.2 & \bf 99.6 & 93.4 \\
%   & \gc \texttt{Qwen3-8B} + SFT & \gc \textbf{20.4}\,\dup{7.6} & \gc \textbf{57.6}\,\dup{15.4} & \gc 99.2\,\ddn{0.4} & \gc \textbf{97.8}\,\dup{4.4} \\

\midrule
\multirow{5}{*}{\shortstack[l]{\texttt{GPT-OSS-120B}\\(medium reasoning)}}
  & No Critic          & 20.4 & 57.4 & 95.2 & 94.2 \\
  & \texttt{Qwen3-4B}  & 16.8\,\ddn{3.6} & 46.6\,\ddn{10.8} & \textbf{100.0}\,\dup{4.8} & \textbf{99.2}\,\dup{5.0} \\
  & \gc + SFT          & \gc 31.2\,\dup{10.8} & \gc 75.2\,\dup{17.8} & \gc 99.8\,\dup{4.6} & \gc 98.0\,\dup{3.8} \\
  & \gc + SFT + DPO    & \gc \textbf{34.8}\,\dup{14.4} & \gc \textbf{75.8}\,\dup{18.4} & \gc 99.2\,\dup{4.0} & \gc 97.2\,\dup{3.0} \\
  & \gc \texttt{Qwen3-8B} + SFT & \gc 31.4\,\dup{11.0} & \gc 74.8\,\dup{17.4} & \gc 99.2\,\dup{4.0} & \gc 96.8\,\dup{2.6} \\

\midrule
\multirow{5}{*}{\shortstack[l]{\texttt{o3-mini}\\(200B)}}
  & No Critic          & 19.0 & 69.4 & 99.4 & \bf 100.0 \\
  & \texttt{Qwen3-4B}  & 20.4\,\dup{1.4} & 64.4\,\ddn{5.0} & \textbf{100.0}\,\dup{0.6} & \textbf{100.0}\,\dzero \\
  & \gc + SFT          & \gc 27.6\,\dup{8.6} & \gc 73.2\,\dup{3.8} & \gc \textbf{100.0}\,\dup{0.6} & \gc 99.6\,\ddn{0.4} \\
  & \gc + SFT + DPO    & \gc 28.2\,\dup{9.2} & \gc \textbf{75.2}\,\dup{5.8} & \gc \textbf{100.0}\,\dup{0.6} & \gc 99.4\,\ddn{0.6} \\
  & \gc \texttt{Qwen3-8B} + SFT & \gc \textbf{29.4}\,\dup{10.4} & \gc 71.0\,\dup{1.6} & \gc \textbf{100.0}\,\dup{0.6} & \gc 98.8\,\ddn{1.2} \\
\bottomrule
\end{tabular}%
}
\caption{Results on SWE-Bench Verified. We report the better result from $k\in\{5,10\}$ for every configuration. \emph{Non-Loop \%} is the fraction of trajectories that do \emph{not} repeat a Bash command in three consecutive steps. ``+ SFT'' is a critic trained on teacher critiques generated with the high-level prompt (a \cwm{} + \qwenNext{} mix). ``+ SFT + DPO'' further optimizes the 4B SFT critic with DPO on critique preference pairs. All arrows compare to ``No Critic" setting. \textbf{Bold} highlights the best non-Opus value in each column.}
\label{tab:main}
\end{table*}

\start{Evaluation Metrics}
% \label{sec:setup_metrics}
We report evaluation metrics on both the output quality and the cost: 
(i) \underline{Resolve Rate}, the fraction of the 500 instances whose submitted patch passes both the \texttt{FAIL\_TO\_PASS} and \texttt{PASS\_TO\_PASS} test sets; 
% (ii) \textbf{resolve-on-submitted rate}, resolve rate conditioned on submission, isolating patch quality from any change in submission rate; 
(ii) \underline{Localized Rate}, the percentage of evaluated instances whose patch edits at least one file also touched by the gold patch~\citep{xia2025agentless}.
(iii) \underline{Non-Empty Rate}, the fraction of runs that exit through the agent's submit command within the $B{=}150$ budget. 
(iv) \underline{Non-Stuck-in-Loop Rate}, the percentage of trajectories that do not contain the same command in three consecutive assistant steps.
(v) \underline{Total Cost per Instance}, the mean combined agent and critic cost in USD per instance, computed from token usage and the rates in App.~\ref{app:pricing}.

\start{Coding Agents} 
We evaluate the critic model with six different coding agents: \textbf{\qwenThirtyTwoB{}}, \textbf{\qwenNextEightyB{}}~\citep{qwen3}, 
\textbf{\texttt{GLM-4.7-Flash}}~\citep{glm47flash}, \textbf{\texttt{GPT-OSS-20B}},
\textbf{\texttt{GPT-OSS-120B}}~\citep{openai2025gptoss}, and \textbf{\texttt{o3-mini}}~\citep{openai2025o3mini}. 
The coding agents are frozen throughout. 
We use the same critic model checkpoint for all coding agents.

\start{Critic Models} We first use a strong proprietary LLM as the critic model, \textbf{\opus{}} \citep{claude46}, which is also used as the teacher model in the SFT stage and the LLM-judge in the DPO stage.
Then we evaluate our trained critic models. 
We conduct SFT training on both \textbf{\texttt{Qwen3-4B}} and \textbf{\qwenEightB{}}~\citep{qwen3}. We conduct SFT+DPO training only on the 4B model due to constraints on computational resources.
We additionally evaluate the untrained 4B and 8B models as the critic models.
The baseline is denoted as ``No critic'', where the coding agents complete the task without any guidance.
We design and compare with other ablations in \S\ref{sec:detailed_analyses}.

% \subsection{Critic Configurations}
% \label{sec:setup_conditions}
% We evaluate no-critic and trained-critic conditions on all six agents. On the two Qwen agents, we additionally compare untrained off-the-shelf critics and the \opus{} teacher. Each prompted critic uses one of two prompt formats, \textbf{detailed} or \textbf{high-level} (App.~\ref{app:critic_prompts}).

% \start{(1) No critic} The unguided agent.

% \start{(2) Untrained critics} Off-the-shelf Qwen3-4B or \qwenEightB{}.

% \start{(3) SFT-trained critics} Qwen3-4B or \qwenEightB{} trained on \opus{}-generated critiques (§\ref{sec:sft}).

% \start{(4) Teacher} \opus{}.

\subsection{Implementation Details}
\label{sec:details}
\start{Inference Details}
All agent uses the mini-SWE-agent scaffold~\cite{swe-agent}, with a budget of $150$ steps, $\$6.0$, and $60$-seconds per-command.
To reduce variance, every critic-guided run shares its first $k$ agent steps with the "No critic" run. 
This is based on our observation that differences in early steps could cascade into divergent trajectories.
% This isolates the critic's contribution from rollout variance: small differences in early actions otherwise cascade into divergent observations (e.g., reading lines $50$-$100$ instead of $100$-$150$ of a file).

\start{Training Details}
We collect training trajectories from $500$ instances randomly sampled from R2E-Gym tasks in SWE-Bench repositories (disjoint from SWE-Bench Verified at the instance level).
We collect training corpora using the combination of \cwm{} and \qwenNext{} as the coding agents.
Additional details can be found in \autoref{app:details}.

% \section{Results and Analysis}
% \label{sec:results}

% Helper macros for rate-with-delta cells.
% \uparr: positive delta (green), \downarr: negative delta (red), \nodelta: no comparison
\newcommand{\uparr}[1]{\,\textcolor{darkgreen}{\scriptsize($\uparrow$#1)}}
\newcommand{\downarr}[1]{\,\textcolor{red}{\scriptsize($\downarrow$#1)}}

\subsection{Main Results}
\label{sec:main_results}

Table~\ref{tab:main} reports our main results on SWE-Bench Verified across six agent models. 
% The trained critic is always queried with the high-level prompt at inference: it is smaller than the agent, and the high-level prompt prevents it from backseat-driving with action-level prescriptions it is not qualified to give (§\ref{sec:critic_models}). 
% Our SFT method in Table~\ref{tab:main} uses \texttt{Qwen3-4B} or \qwenEightB{}, trained on high-level teacher critiques (a \cwm{} + \qwenNext{} mix). For the two Qwen agents, we also train an 8B critic on detailed teacher critiques and compare it against the high-level critic directly (Table~\ref{tab:prompt_ablation}).

% \paragraph{How well does the trained critic perform?} 
\start{Our Trained Critic Models Significantly Improve All Six Coding Agents}
Across the six agent models in \autoref{tab:main}, our critic model improves the resolved rate over the ``No Critic'' baseline by $5.6\%$--$16.0\%$ absolute gain. All six gains are statistically significant (App.~\ref{app:significance}). 
We can also see that the SFT+DPO method achieves better performance than the SFT-only method for all six coding agents and the SFT-only method always outperforms the untrained critic model and the ``No Critic'' baseline, which demonstrates the effectiveness of both the SFT and the DPO training stages.
Namely, the performance gain mainly comes from training rather than merely interleaving another model into the agent's loop.

% \paragraph{Is the gain from training, or from interleaving any 8B model?} 
\start{Training Improves Performance Across Different Critic Model Sizes}
Relative to the matched untrained 4B critic, SFT improves the resolved rate for all six coding agents by $1.2\%$--$14.4\%$, and DPO adds additional gain of $0.6\%$--$5.0\%$.
Compared to the untrained 8B critic, SFT improves the resolved rate for the two Qwen agents by $3.8\%$ and $4.6\%$, respectively. 
The results show that our training data is effective across different critic model sizes.

\subsection{Detailed Ablations and Analyses}
\label{sec:detailed_analyses}
\start{Critic Models Transfer across Programming Languages}
As shown in \autoref{tab:multilingual}, on Multi-SWE-Bench, the 8B SFT critic model improves the resolved rate by $2.0$ points on both \qwenNext{} and \qwenThirtyTwoB{}. Its SFT data contain only Python trajectories, whereas this 300-instance benchmark spans nine non-Python languages, showing that the trained critic model's guidance transfers beyond its training language distribution.

\begin{table}[t]
\centering
% \small
% \setlength{\tabcolsep}{4pt}
\resizebox{0.95\linewidth}{!}{%
\begin{tabular}{@{}l l c c@{}}
\toprule
\textbf{Size} & \textbf{Critic Model} & \textbf{\texttt{Qwen3-Next}} & \textbf{\texttt{Qwen3-32B}} \\
\midrule
\multirow{2}{*}{8B} & No Critic & 9.7 & 1.3 \\
& + SFT & 11.7\,\dup{2.0} & \textbf{3.3}\,\dup{2.0} \\
\bottomrule
\end{tabular}%
}
% \begin{tabular}{@{}lcc@{}}
% \toprule
% \textbf{Agent} & \textbf{No critic} & \textbf{+ SFT} \\
% \midrule
% \qwenNext{} & 9.7 & \textbf{11.7}\,\dup{2.0} \\
% \qwenThirtyTwoB{} & 1.3 & \textbf{3.3}\,\dup{2.0} \\
% \bottomrule
% \end{tabular}
\caption{Resolve rate (\%) on 300 Multi-SWE-Bench instances. The 8B CWM+Qwen critic uses $k{=}5$ and is trained only on Python trajectories. Arrows show gains over the ``No Critic'' baseline.}
\label{tab:multilingual}
\end{table}

\start{Critic Models Transfer Across Different Coding Agents in Training and Evaluation}
\autoref{tab:critic_ablation} varies the SFT corpus, where we generate critiques on the trajectories generated by the \cwm{} model only (denoted as ``CWM-only''), the \qwenEightB{} model only (``Qwen-only''), and our default SFT corpus that combines both coding agents (``Qwen+CWM'').
We can see that the 8B critic model trained on CWM trajectories alone also improves \qwenNext{} and \qwenThirtyTwoB{} by $+3.6$ and $+2.6$ points, respectively, showing that the learned steering signal transfers beyond coding agents in training and evaluation.
Furthermore, \qwenNext{} favors Qwen-only training at both critic sizes, while Qwen+CWM training produces more balanced gains across the two agents (e.g., with an average gain of 3.4\% for the 4B critic model and 5.1\% for the 8B model).
This indicates that training on the corpora with multiple coding agents yields better overall gain across diverse coding agents.

\begin{table}[t]
\centering
% \small
% \setlength{\tabcolsep}{3pt}
\resizebox{0.90\linewidth}{!}{%
\begin{tabular}{@{}l l c c@{}}
\toprule
\textbf{Size} & \textbf{Corpus} & \textbf{\texttt{Qwen3-Next}} & \textbf{\texttt{Qwen3-32B}} \\
\midrule
-- & -- & 20.0 & 8.8 \\
\midrule
\multirow{2}{*}{4B} 
& Qwen-only & \textbf{24.6}\,\dup{4.6} & 10.8\,\dup{2.0} \\
& Qwen+CWM & 24.2\,\dup{4.2} & \textbf{11.4}\,\dup{2.6} \\
\midrule
\multirow{3}{*}{8B} 
& CWM-only & 23.6\,\dup{3.6} & 11.4\,\dup{2.6} \\
& Qwen-only & \textbf{26.2}\,\dup{6.2} & 10.6\,\dup{1.8} \\
& Qwen+CWM & 25.2\,\dup{5.2} & \textbf{13.8}\,\dup{5.0} \\
\bottomrule
\end{tabular}%
}
\caption{Resolved rate (\%) of critic models trained with different SFT corpus on SWE-Bench Verified. All critic models use $k{=}5$; arrows show gains over no critic.}
\label{tab:critic_ablation}
\end{table}

\begin{figure}[!t]
\centering
\includegraphics[width=\linewidth]{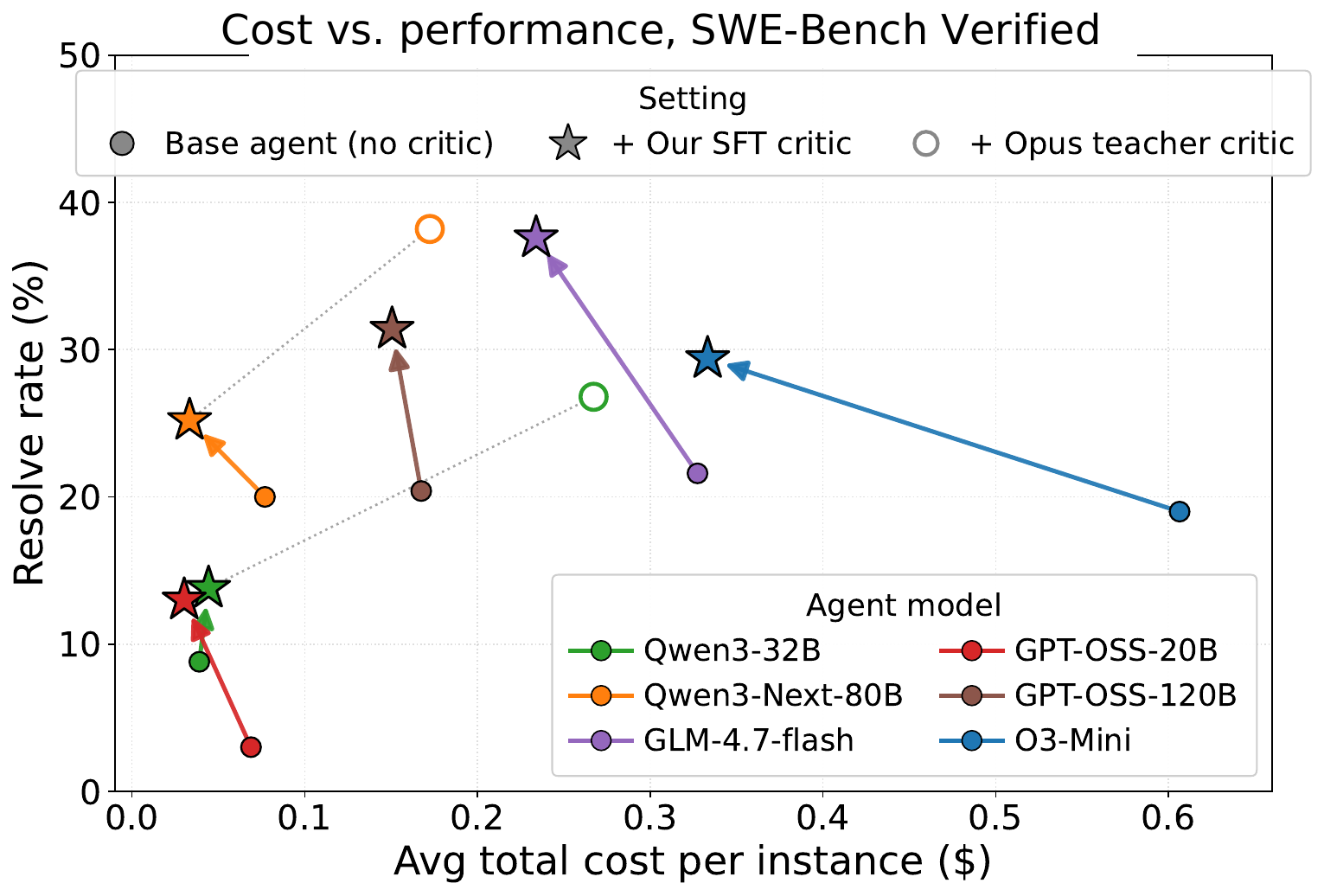}
\caption{Cost versus performance on SWE-Bench Verified across six coding agents. Each agent shows its base (\ding{108}) and the 8B SFT critic ($\bigstar$); the two Qwen agents additionally show the \opus{} teacher critic ($\circ$). The cost is for the agent \textbf{plus} the critic model.}
\label{fig:pareto_full500}
\end{figure}

% \paragraph{What does the gain cost relative to the teacher critic?} 
\start{Our Trained Critic Models Achieve Balance between the Coding Agents' Performance and Cost}
As shown in \autoref{fig:pareto_full500}, on the two Qwen agents, the entire critic-guided system costs \$0.03--\$0.04 per instance with the 8B SFT critic model, versus \$0.17--\$0.27 with \opus{}.
Namely, our trained critic model achieves a $5.2$--$6.0\times$ reduction in total cost.
% Total per-instance cost is $4$-$5\times$ below \opus{} on every agent model. 
Specifically, on \qwenNext{}, the 8B SFT critic model Pareto-dominates the ``No Critic'' baseline, which is strictly better on both resolved rate ($25.2\%$ vs.\ $20.0\%$) and total per-instance cost (\$0.03 vs.\ \$0.08).
The reason is that with the critic model, the \qwenNext{} coding agent requires on average fewer steps to resolve a task, which reduces the overall inference cost.

% \paragraph{Which prompt format produces the better trained critic?} 
\start{Having High-level Critique Only Improves Training Performance}
% The high-level one, even though the detailed prompt wins at the teacher.
We compare to an ablation called ``Detailed-Prompt'' in \autoref{tab:opus_prompt_ablation}. When we generate SFT data, we do NOT require the teacher critic model to only generate high-level guidance.
Namely, the SFT critiques contain both high-level guidance and low-level implementation details such as the concrete tool call commands.
Results show that on both Qwen agents, high-level supervision outperforms detailed supervision: $13.8$ vs.\ $12.6\%$ on \qwenThirtyTwoB{} and $25.2$ vs.\ $24.6\%$ on \qwenNext{}. 
This result is consistent with the hypothesis in §\ref{sec:critique_design}, that the \opus{} teacher's detailed critiques may contain low-level implementation knowledge that the Qwen agents could in principle exploit, but the smaller student critic model cannot reliably learn. The high-level prompt (i.e., our design choice) instead restricts supervision to signals that the trained critic model can more faithfully inherit.

% Prompt-format ablation: an 8B critic trained on detailed-prompt vs
% high-level-prompt teacher critiques, evaluated on the two Qwen agent models.
% Resolve rate only; same numbers as the corresponding rows of Table~\ref{tab:main}.
\begin{table}[t]
\centering
\resizebox{\linewidth}{!}{%
\begin{tabular}{@{}l l c c@{}}
\toprule
\textbf{Size} & \textbf{Critic Model} & \textbf{\texttt{Qwen3-Next}} & \textbf{\texttt{Qwen3-32B}} \\
\midrule
\multirow{3}{*}{8B} & No Critic & 20.0 & 8.8 \\
& + SFT (Detailed-Prompt) & 24.6\,\dup{4.6} & 12.6\,\dup{3.8} \\
& + SFT (High-Level, Ours) & \textbf{25.2}\,\dup{5.2} & \textbf{13.8}\,\dup{5.0} \\
\bottomrule
\end{tabular}%
}
% \small
% \setlength{\tabcolsep}{4pt}
% \begin{tabular}{@{}lcc@{}}
% \toprule
% \textbf{Agent} & \textbf{Detailed} & \textbf{High-Level (Ours)} \\
% \midrule
% \qwenThirtyTwoB{} & 12.6 & \textbf{13.8} \\
% \qwenNext{} & 24.6 & \textbf{25.2} \\
% \bottomrule
% \end{tabular}
\caption{Resolve rate (\%) on SWE-Bench Verified for 8B critic models trained on SFT data with the detailed critiques vs.\ the high-level teacher critiques (our data).}
\label{tab:prompt_ablation}
\end{table}

\subsection{Qualitative Analysis and Case Studies}
\label{sec:qualitative}
We conduct qualitative analyses with concrete cases to answer the following research questions: 
(1) What does the critic model learn from training? And
(2) The ``Detailed-Prompt'' ablation achieves better performance than our ``High-level'' prompt for the teacher critic model \opus{}, is this expected?
% §\ref{sec:main_results} leaves two follow-ups open. 
% \emph{What does training actually change in the critic's outputs}, given that an off-the-shelf 8B critic improves resolved rate by at most $1.2$ points? 
% And \emph{why does the detailed prompt win at the \opus{} teacher but lose at the 8B trained critic}, reversing the prompt-format ranking? 
% All case studies are drawn from base, untrained-critic, and trained-critic runs that diverge in outcome, plus a 50-instance \opus{} teacher subsample run under both prompt formats.

\start{What Critic Model Training Removes, and What it Adds} 
As analyzed in App. \ref{app:untrained_vs_trained}, the untrained \qwenEightB{} critic model fails in two ways: only $43.4\%$ of its outputs correctly follow the format in the prompt (vs.\ $85.8\%$ after SFT) and $19.3\%$ contain hallucinated bash command actions that are not grounded in the coding agent's trajectory (vs.\ $0\%$ after SFT). 
In addition, when the untrained critic model does raise an error, it almost always flags only Verification Failures ($330/354$ firings) with a ``run the test script'' recovery, pulling the agent off self-review even on instances the base agent would otherwise have salvaged. 

We also check the evaluation instances where the trained critic model helps the coding agent succeed, but the coding agent alone cannot resolve. 
The common pattern is a strategy-level redirect: the critic identifies the failure category the agent is in and points to a different class of approach, without dictating the specific code. 
We provide two case studies that illustrate the pattern in App.~\ref{app:untrained_vs_trained} and App.~\ref{app:sandbox_adaptation}. 
On instance \texttt{django\_\_django-9296}, \texttt{sed} append commands silently duplicate a method each time the pattern matches. The critic jointly flags Step Repetition and Tool Selection Errors and redirects the agent to a heredoc rewrite, which resolves the task. 
On \texttt{sympy\_\_sympy-22714}, a \texttt{ModuleNotFoundError: mpmath} blocks execution-based verification; the critic names the import as the root cause and tells the agent to bypass it and read the source directly.

% \paragraph{Backseat Driving in the Detailed Teacher Prompt.} 
\start{Detailed Critiques Work Better than High-level Ones for Strong Teacher Models Because of Backseat Driving}
We show in \autoref{tab:opus_prompt_ablation} that the teacher critic model performs overall better when we allow it to generate detailed critiques instead of requiring it to only generate high-level guidance (e.g., \qwenThirtyTwoB{} achieves a resolved rate of 40.6\% when combined with detailed critiques, but only achieves 26.8\% with high-level critiques).

We manually examine the examples and observe that if the critic model suggests a concrete action (e.g., the tool call command),
the coding agent is likely to directly copy it in its next action.
\autoref{tab:critic_share} presents an analyses on a 50-instance subsample with the \opus{} teacher. 
If we use the ``detailed'' prompt for the critic model, for 23 out of 50 instances, more than 30\% of the lines in the final submitted code patch are directly copied from the critic model.
In comparison, when using the high-level prompt, the coding agent copies more than 30\% of the code lines in only 1/50 of the evaluation instances.
The detailed prompt also increases the average number of lines in the generated critiques (e.g., from $1.3$ to $6.7$ for the $k=5$ experiments).
% Let $\rho$ denote the fraction of the agent's final-submission lines copied verbatim from the critic. 
% On a 50-instance subsample with the \opus{} teacher, the detailed prompt emits $6.7$ code lines per firing and yields $\rho \geq 0.30$ on $23/50$ instances at $k{=}5$; the high-level prompt cuts these to $1.3$ and $1/50$ (Table~\ref{tab:critic_share}). 
The results suggest that the detailed critiques' higher resolved rate ($32/50$ vs.\ $22/50$) is driven by direct copying rather than agent reasoning: when the teacher dictates a single patch, the agent transcribes it and resolves; when the teacher offers multiple candidate fixes without committing, the agent picks the wrong action and writes a confidently incorrect patch. 

While the strong proprietary critic model such as \opus{} has the capability to generate correct actions, the student critic model has a much smaller model size, even much smaller than the coding agents.
Therefore, it is unlikely that the distilled 8B critic model is strong at generating both high-level guidance and concrete actions, and training with high-level critiques only yields better overall performance.

\begin{table}[t]
\centering
% \small
\setlength{\tabcolsep}{4pt}
\resizebox{\linewidth}{!}{%
\begin{tabular}{@{}l c c c c@{}}
\toprule
\textbf{Prompt} & $k$ & \textbf{Avg. Critique Lines} & \textbf{Copy\%} & \textbf{$\geq 0.30$\%} \\
\midrule
Detailed & 5  & 6.7 & 0.32 & 23/50 \\
High-Level & 5  & 1.3 & 0.01 & 1/50  \\
\midrule
Detailed & 10 & 7.1 & 0.27 & 10/49 \\
High-Level & 10 & 1.9 & 0.00 & 0/49  \\
\bottomrule
\end{tabular}
}
\caption{\opus{} critic model backseat driving on the 50 SWE-Bench Verified mini examples. 
\underline{Avg. Critique Lines} denotes the average number of lines in the critiques.
\underline{Copy\%} denotes the fraction of lines in the submitted code patch that are directly copied from the critic model.
\underline{$\geq 0.30$\%} counts the instances where more than 30\% of the coding agent’s final submission lines are directly copied from the critic model.
We test critique frequencies of $k=5$ and $10$.
}
\label{tab:critic_share}
\end{table}

% \subsection{Effect of Agent Backbone}
% \label{sec:ablation_agent}

% To check whether the critic gain depends on the agent's RL post-training, we replace the RL-trained CWM-32B with the same model after SFT only. The SFT-only agent resolves 66/500 (13.2\%) without a critic; pairing it with our trained critic on a partial run did not close the gap to the RL-trained baseline. We report this as evidence that the critic complements RL post-training rather than substituting for it.

\section{Related Work}

\start{Critic and Advisor Models for Generation.}
A growing body of work studies \emph{critic} or \emph{advisor} models that improve generation through evaluative feedback.
\citet{gou2023critic} established critique as an intermediate signal for improving model outputs.
In code and math, \citet{mcaleese2024llm, gao2025llm} train critics that give natural-language feedback on model-written solutions, \citet{yadavally2025large} evaluates code edits without execution, and \citet{wang2026rubric} shows that effective critics can be learned from sparse outcome signals using rubric-based supervision.
Outside code, \citet{xie-etal-2025-factuality} trains FenCE, a claim-level critic for factuality, while \citet{asawa2026trainadvisorsteeringblackbox} learns small advisor models that steer stronger black-box models via instance-specific instructions.
These critics operate on completed outputs or through the agent's initial prompt, without revisiting the agent once generation is underway.
Issue resolution instead involves long, tool-using trajectories where many failures stem from flawed reasoning rather than incorrect low-level actions~\citep{liu2025empiricalfailures}.

We therefore train a compact critic that intervenes at intermediate steps to provide trajectory-aware, strategy-level feedback on \emph{how} a frozen coding agent should proceed, rather than judging only the final artifact.
% This trades additional inference-time compute (one critic call every $k$ agent steps) for the ability to redirect the agent before it commits to a wrong path.
% We show in \S\ref{sec:experiments} that the critic improves resolve rate while having significantly lower total cost than proprietary models.

\start{Auxiliary Models for Coding Agents.}
Prior work improves coding agents by altering the agent's context, architecture, or parameters.
For example, \citet{wang2026swe} improves issue resolution via context selection and pruning. Modular systems instead decompose software-engineering agents into specialized roles such as planners and executors~\citep{arora2024masaimodulararchitecturesoftwareengineering,10.5555/3780338.3780932}, and post-training methods directly optimize the coding agent itself~\citep{pmlr-v267-pan25g,yang2025swesmith,xie2026hybridgymtrainingcodingagents,jain2025regym}.
These approaches still couple strategy-level reasoning with the agent's implementation behavior (tool calls, code edits, prompt-specific action formats).

We instead keep the agent frozen and train the critic model with a more specific role: it only supplies high-level feedback while the agent remains responsible for step implementation.
This specific role makes the critic model easier to train from existing training environments, more likely to capture transferable reasoning patterns, and applicable to agents without parameter access.
\section{Conclusion and Future Works}
\label{sec:conclusion}

We study whether a small, trained critic model can provide helpful strategy-level signals for a larger coding agent. 
We construct SFT and DPO training data using a strong proprietary model (\opus{}) as the teacher model and the LLM judge.
Specifically, for the SFT stage, we design the structure of the critiques, which includes error detection for the coding agent, high-level feedback, and a budget alert.
We also design preference data for the DPO stage, focusing on the correctness and clarity of the guidance to future steps.
Experiments show that our trained critic models significantly improve six coding agents by $5.6\%$--$16.0\%$ resolved rates on SWE-Bench Verified.
We also demonstrate that our training data transfers across programming languages and coding agents and works for critic models with different sizes.
Our detailed analyses further demonstrate the effectiveness of both the SFT and DPO stages and verify the important design choice of training on high-level critiques only. 

\start{Future Works}
We encourage future researchers to keep working on the direction of critic model training.
Specifically, while we observe performance gains in the resolved rate for all coding agents, we do not always improve intermediate metrics such as the non-empty rate or the non-loop rate, and we leave it to future work to have a deep analysis on this phenomenon.
Furthermore, in this work, we restrict the frequency of giving critiques to every $k=5$ or $10$ steps. However, it is possible to train a critic model to dynamically select the step to provide feedback.
Finally, the usage of critic models is just one example of multi-model collaboration frameworks in inference.
We hope there could be more research that studies the inference settings involving multiple agents or models, with the aim of either improving the overall performance or reducing the inference cost.

% restricts them to high-level feedback, and intervenes every $k$ steps with budget-aware instructions. On SWE-Bench Verified, the best trained configuration improves all six agents by $5.6$--$16.0$ points. A critic trained on \cwm{} alone already transfers to unseen agents ($2.6$--$3.6$ points), and mixing Qwen trajectories into the corpus increases those gains to $5.0$--$5.2$ points at $5.2$--$6.0\times$ lower total cost than the teacher. On \qwenNext{}, the critic-guided system is also cheaper than the unguided agent. These transfer gains, together with the smaller $0.2$--$1.4$-point gains from same-size untrained controls on those agents, point to the supervision rather than a second model in the loop as the source of improvement. Our results show that training critic models is a practical complement to scaling agent training.

% \newpage
\section*{Limitations and Potential Risks}
We identify the limitations of this work as follows:

\start{Evaluation Settings}
In our experiments, we only conduct evaluation on the SWE-Bench Verified and Multi-SWE-Bench benchmarks under the mini-SWE-agent scaffold, with a 150-step budget and periodic critic intervention. In principle, the critic models can also be applied to other coding benchmarks and other agent scaffolds.

\start{Error Taxonomy} 
The error taxonomy was also developed specifically for the issue-solving task. We may need to develop new error taxonomies for different coding tasks.

\start{Use of Teacher Models and LLM Judges}
Our critic models inherit limitations of the teacher critic model's supervision, and the SFT+DPO variants may additionally depend on the preference judge. 

\start{Potential Risks}
We do not identify risks specific to this work beyond those generally associated with the use of large language models.

% Due to compute limitations, we do not show results over multiple runs per configuration. However, we report statistical significance in Appendix~\ref{app:significance}. Critic interventions can also lengthen trajectories or induce repeated actions; if an agent exhausts the 150-step budget without submitting a patch, the instance is automatically unresolved. Future work should optimize critic training for downstream agent outcomes and trajectory efficiency to reduce these budget-exhaustion failures. 

\vspace{0.15cm}
\section*{Acknowledgments}
We would like to thank Daniel Fried for their feedback on the development of the DPO training stage. We thank Graham Neubig and Dan Roth for providing overall feedback on this work.

This work was supported in part by NSF grant DSES-2222762. 

\newpage
\bibliography{custom}

\clearpage
\appendix

\section{Analyses on the Gap between Teacher and Student Critic Models}
\label{app:analyses}
\begin{figure*}[t]
    \centering
    \includegraphics[width=\linewidth]{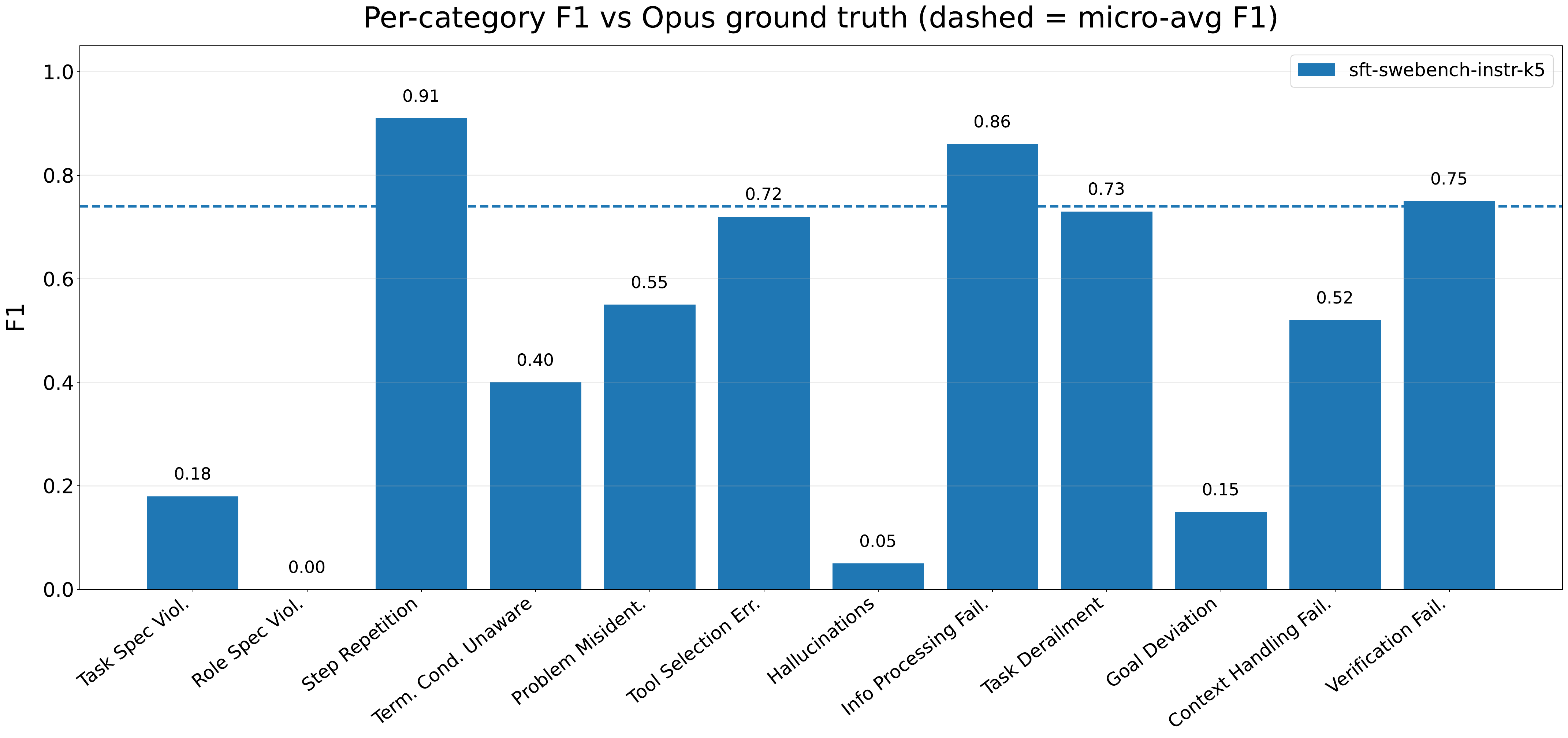}
    \caption{Error categorization results.}
    \label{fig:analysis_error_cat}
\end{figure*}

\start{Error Categorization Results}
We first compute the agreement of error categorization between the student and teacher critic models. As shown in \autoref{fig:analysis_error_cat}, the agreement rate is high for major error categories such as step repetition (F1=$0.91$) and task derailment (F1=$0.73$).
The overall average Micro-F1 score is $0.74$.

\begin{figure*}[t]
    \centering
    \includegraphics[width=\linewidth]{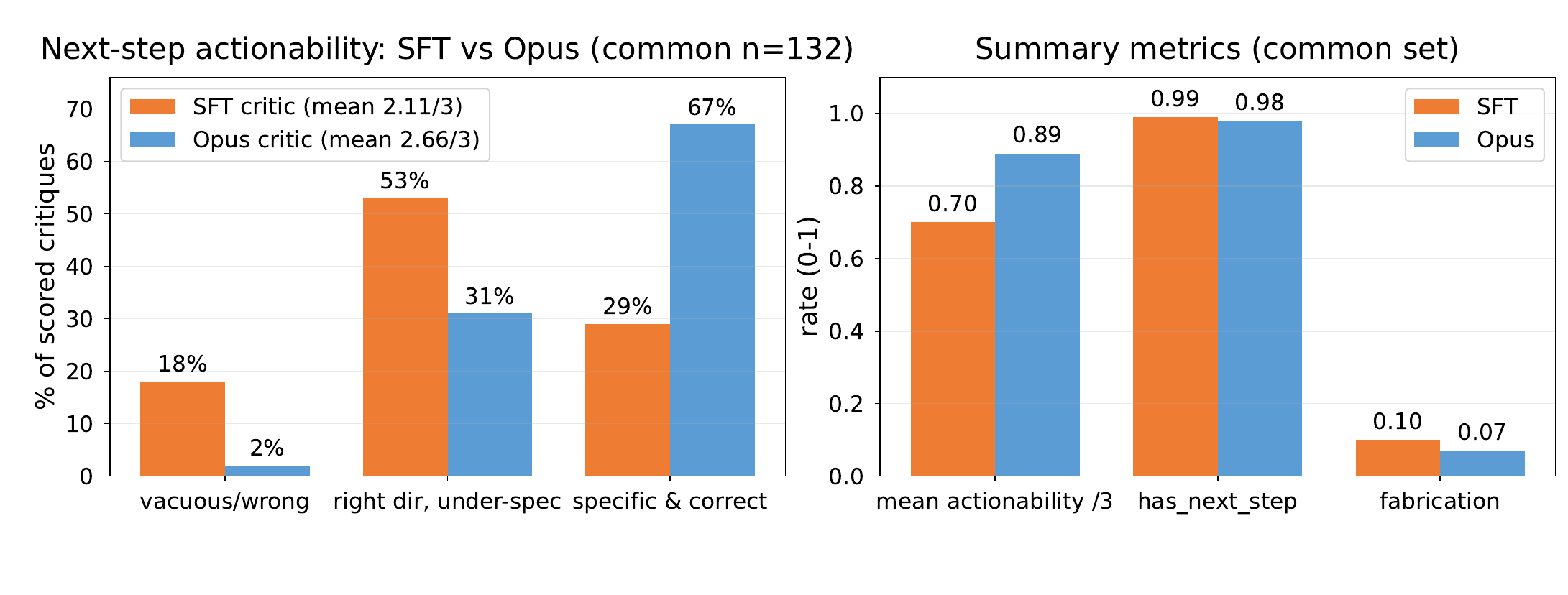}
    \caption{Guidance quality analysis.}
    \label{fig:analysis_guidance}
\end{figure*}

\start{Overall Guidance Quality}
Then we prompt an LLM judge to check whether the guidance generated by both the teacher and student models is both correct and well-specified.
We use a different model (\texttt{DeepSeek V4 Pro}) as the judge to avoid self-bias.
The results are shown in \autoref{fig:analysis_guidance}, the SFT critic model has a significantly lower rate of generating correct and well-specified guidance compared to the teacher critic model.

\section{Training and Inference Details}
\label{app:details}

\subsection{Inference Details}
\label{sec:setup_impl}

\start{Scaffold} All agent runs use the mini-SWE-agent scaffold~\cite{swe-agent}, with a budget of $150$ steps, $\$6.0$ and $60$-seconds per-command.

\start{Sampling and serving} Open-weight models are served with vLLM\footnote{\url{https://github.com/vllm-project/vllm}} on $8\times$L40S; \opus{} is accessed via its hosted API. We sample at \texttt{temperature} $0$, \texttt{top-p} $1$, with a per-turn output cap of $2048$ tokens. For \texttt{GPT-OSS}, we use \texttt{medium} reasoning effort.

\start{Shared prefix} Every critic-guided run shares its first $k$ agent steps with the base run. This isolates the critic's contribution from rollout variance: small differences in early actions otherwise cascade into divergent observations (e.g., reading lines $50$-$100$ instead of $100$-$150$ of a file).

\start{Default settings} Our trained critic uses the \textbf{high-level} prompt and budget-aware instructions (§\ref{sec:critique_design}).

\subsection{Training Details}
\label{sec:setup_training}
% \start{Critic training} 
Training trajectories are collected on $500$ instances randomly sampled from R2E-Gym tasks in SWE-Bench repositories (disjoint from SWE-Bench at the instance level), querying \opus{} every $k=5$ agent steps for trajectory feedback. We collect two corpora: a \cwm{}-only corpus, and a mixed corpus that adds \qwenNext{} trajectories on top of it. The mixed corpus is used for the main trained critics; the \cwm{}-only corpus is used for the transfer experiment (§\ref{sec:main_results}). We denoise the data by discarding teacher responses that lack the expected error-taxonomy markers (likely format failures or agent text mistakenly logged as feedback). To fit the $32{,}768$-token context, we drop the oldest trajectory and feedback blocks first while preserving the system prompt, task description, final prompt, and teacher response in full. The \cwm{}-only corpus yields $4{,}532$ training samples under the high-level prompt and $3{,}135$ under the detailed prompt; the mixed corpus adds a comparable shard of \qwenNext{} samples. We train \texttt{Qwen3-4B} and \qwenEightB{} for 3 epochs with AdamW (peak LR $5{\times}10^{-6}$, cosine schedule, $10\%$ linear warmup, bfloat16); loss is computed only on teacher critique tokens, with agent actions, observations, and prior critiques kept in context but masked out. Global batch size is 8 across $8{\times}$L40S with gradient checkpointing.

\section{Additional Related Work}
\paragraph{Textual feedback for prompt optimization.}
A related line of work uses textual feedback not as the output of a standalone critic model, but as a signal for improving prompts, roles, or coordination strategies within a larger agent system. 
\citet{shen2025optimizing} is a representative example, showing that failure feedback can be used to improve a multi-agent coding system. 
Likewise, \citet{pandita2025prorefine} studies refinement of prompts using feedback signals at inference time. 
Closely related, \citet{shinn2023reflexion} proposes to improve agents via natural language reflections stored in memory, enabling agents to learn from prior failures without parameter updates. 
These approaches demonstrate that textual feedback can serve as a powerful learning signal. 
However, their primary optimization target is the prompting strategy or in-context behavior of the agent, rather than a compact critic model that can serve as an independent advisor during issue resolution.
\section{Method Preliminaries}
\label{app:method_prelim}

\paragraph{Division of labor.} The coding agent $\pi_A$ handles concrete repository interaction: inspecting files, running commands, editing code, and deciding when to submit. The critic instead provides high-level judgments about the \emph{trajectory}: whether the agent is making progress, pursuing an unsupported assumption, repeating an unproductive strategy, or needs to reconsider its overall direction.

\paragraph{Why high-level feedback.} We do not want the critic to become a second actor that ``backseat drives'' $\pi_A$ with commands, file-level instructions, or step-by-step recovery plans. The coding agent is already the more capable model for concrete tool use and code editing; asking a smaller critic to prescribe specific actions risks replacing its reasoning with weaker instructions. We therefore train the critic to provide \emph{high-level strategic feedback}, not action-level control.

\paragraph{Scaffold details.} The scaffold gives the coding agent a simple command-line interface rather than a large set of specialized tools. Keeping the scaffold minimal lets us study the effect of the critic without confounding it with scaffold-specific machinery. Each task instance $x$ consists of a natural-language issue description and a repository snapshot. At each step, the coding agent $\pi_A$ may reason about the task, execute shell commands, inspect files, edit code, or submit a final patch. A run succeeds if the submitted patch passes the benchmark's evaluation harness.

\section{Critic Prompts}
\label{app:critic_prompts}

This appendix gives the two critic prompts compared in §\ref{sec:critique_design} and §\ref{sec:main_results}: the \emph{high-level} prompt that constrains the critic to high-level feedback, and the \emph{detailed} prompt that permits action-level prescriptions. Both prompts adapt the trajectory-feedback prompt of \citet{gandhi2025sweprm} and share its 12-category trajectory-level error taxonomy. The detailed prompt uses the upstream version unchanged; the high-level prompt adds the framing rules listed in §\ref{sec:critique_design}, removes the per-category recovery hints, and tightens the example response. Both critics receive the same agent-side context (task description, trajectory prefix, and prior critic feedback history) at every $k$ steps. Once the agent passes step $100$ of its $150$-step budget, the budget-aware instructions in §\ref{app:step_aware} are appended to whichever prompt is in use; they prioritize recommending submission as the budget approaches.

\subsection{High-Level Prompt}
\label{app:high_level_prompt}

\begin{lstlisting}[basicstyle=\ttfamily\scriptsize,xleftmargin=-4pt,xrightmargin=-4pt]
You are a supervisor monitoring an LLM-based code agent. Your role is to detect trajectory-level errors and provide corrective guidance to prevent task failure.
Analyze the provided agent trace for the following error categories. For each detected error, provide specific evidence and a recovery action.

The agent works in an isolated repo; assume no hidden external edits. There is EXACTLY ONE final submission command; it is ALWAYS essential.

IMPORTANT RULES:
- Respond ONLY with analytical feedback in the format below. Never generate bash commands, code blocks, or simulated agent/environment output.
- Be CONSERVATIVE. Default to "On track" unless you see clear evidence of a problem. The agent is competent - let it work. Unnecessary interventions cause the agent to second-guess itself and loop, which is worse than a slow but steady trajectory.
- Use "Critical intervention required" ONLY for: (1) source file is corrupted/broken by a bad edit, or (2) agent has repeated the exact same action 3+ times with no progress. Most other issues are "Needs correction" at worst.
- NEVER suggest specific technical approaches, code changes, or where in the code to look. The agent understands the codebase better than you. If the agent's approach seems wrong, say so, but do NOT suggest an alternative - let the agent decide.
- If the agent has applied a fix and verified it works, tell it to submit immediately.
- Keep your response SHORT. 1-2 sentences per detected error. 1-2 sentences for OVERALL_GUIDANCE. Long feedback is counterproductive.

SPECIFICATION ERRORS:

1. Task Specification Violations
Definition: Agent fails to adhere to task constraints or requirements

2. Role Specification Violations
Definition: Agent behaves outside its defined role/responsibilities

3. Step Repetition
Definition: Unnecessary repetition of completed steps or actions

4. Termination Condition Unawareness
Definition: Agent continues working when task completion criteria are met

REASONING ERRORS:

5. Problem Misidentification
Definition: Agent misunderstands the core problem or current subtask

6. Tool Selection Errors
Definition: Agent uses inappropriate tools for the current task

7. Hallucinations
Definition: Agent generates false information or fabricates tool outputs

8. Information Processing Failures
Definition: Poor retrieval of relevant information or misinterpretation

COORDINATION ERRORS:

9. Task Derailment
Definition: Agent deviates from intended objective or loses focus

10. Goal Deviation
Definition: Agent pursues goals that don't serve the main objective

11. Context Handling Failures
Definition: Agent loses important context or forgets previous findings

12. Verification Failures
Definition: Inadequate checking of work quality or correctness

=====================================================
RESPONSE FORMAT
=====================================================

For each error category, respond with:
DETECTED: Yes/No
EVIDENCE: (if detected) One sentence.
RECOVERY_ACTION: (if detected) One sentence. No code, no commands.

Then provide:
TASK_STATUS: Exactly one of: "On track" / "Needs correction" / "Critical intervention required"
OVERALL_GUIDANCE: 1-2 sentences only.

EXAMPLE:

SPECIFICATION ERRORS:
1. Task Specification Violations: DETECTED: No
2. Role Specification Violations: DETECTED: No
3. Step Repetition: DETECTED: Yes
EVIDENCE: Agent ran the same test command three times with identical results.
RECOVERY_ACTION: Stop repeating and analyze the existing test output.
4. Termination Condition Unawareness: DETECTED: No

REASONING ERRORS:
5. Problem Misidentification: DETECTED: No
6. Tool Selection Errors: DETECTED: No
7. Hallucinations: DETECTED: No
8. Information Processing Failures: DETECTED: No

COORDINATION ERRORS:
9. Task Derailment: DETECTED: No
10. Goal Deviation: DETECTED: No
11. Context Handling Failures: DETECTED: No
12. Verification Failures: DETECTED: No

TASK_STATUS: Needs correction
OVERALL_GUIDANCE: You are repeating the same test. Analyze the output you already have and proceed to the next step.

Now review the agent's trajectory and provide your supervisor feedback.
\end{lstlisting}

\subsection{Detailed Prompt}
\label{app:detailed_prompt}

The detailed prompt below is reproduced verbatim from \citet{gandhi2025sweprm} for reproducibility. The high-level prompt above is our modification of it; see §\ref{sec:critique_design} for the added framing rules.

\begin{lstlisting}[basicstyle=\ttfamily\scriptsize,xleftmargin=-4pt,xrightmargin=-4pt]
You are a supervisor monitoring an LLM-based code agent. Your role is to detect trajectory-level errors and provide corrective guidance to prevent task failure.
Analyze the provided agent trace for the following error categories. For each detected error, provide specific evidence and a recovery action.

The agent works in an isolated repo; assume no hidden external edits. There is EXACTLY ONE final submission command; it is ALWAYS essential.

SPECIFICATION ERRORS (System Design Issues)

1. Task Specification Violations
Definition: Agent fails to adhere to task constraints or requirements
Recovery: Redirect agent to original task requirements

2. Role Specification Violations
Definition: Agent behaves outside its defined role/responsibilities
Recovery: Remind agent of its specific role and boundaries

3. Step Repetition
Definition: Unnecessary repetition of completed steps or actions
Recovery: Acknowledge completed work and guide to next logical step

4. Termination Condition Unawareness
Definition: Agent continues working when task completion criteria are met
Recovery: Signal completion criteria and instruct to finalize

REASONING ERRORS (Decision Making Issues)

5. Problem Misidentification
Definition: Agent misunderstands the core problem or current subtask
Recovery: Clarify the actual problem and expected approach

6. Tool Selection Errors
Definition: Agent uses inappropriate tools for the current task
Recovery: Suggest correct tools and explain their appropriate usage

7. Hallucinations
Definition: Agent generates false information or fabricates tool outputs
Recovery: Request verification of claims against actual evidence

8. Information Processing Failures
Definition: Poor retrieval of relevant information or misinterpretation
Recovery: Guide agent to correct information sources and interpretation

COORDINATION ERRORS (Process Management Issues)

9. Task Derailment
Definition: Agent deviates from intended objective or loses focus
Recovery: Realign agent with original objectives and priorities

10. Goal Deviation
Definition: Agent pursues goals that don't serve the main objective
Recovery: Refocus on primary goals and expected outcomes

11. Context Handling Failures
Definition: Agent loses important context or forgets previous findings
Recovery: Provide context summary and key information recap

12. Verification Failures
Definition: Inadequate checking of work quality or correctness
Recovery: Instruct specific verification steps and quality checks

=====================================================
RESPONSE FORMAT
=====================================================

For each error category, respond with:
DETECTED: Yes/No
EVIDENCE: Specific quote or observation from trace (if detected)
RECOVERY_ACTION: Specific instruction to correct the error (if detected)

Then provide:
TASK_STATUS: On track / Needs correction / Critical intervention required
OVERALL_GUIDANCE: Detailed and specific guidance for the agent

Example Response Structure

SPECIFICATION ERRORS:
1. Task Specification Violations: DETECTED: No
2. Role Specification Violations: DETECTED: No
3. Step Repetition: DETECTED: Yes
EVIDENCE: "Agent ran the same test command three times: 'pytest test_file.py'"
RECOVERY_ACTION: "The test has already been executed successfully. Proceed to analyze the results and move to the next development step."
4. Termination Condition Unawareness: DETECTED: No

REASONING ERRORS:
5. Problem Misidentification: DETECTED: No
6. Tool Selection Errors: DETECTED: Yes
EVIDENCE: "Agent used text editor to run Python code instead of using the Python interpreter"
RECOVERY_ACTION: "Use the Python interpreter tool for code execution. The text editor is for viewing and modifying files only."
7. Hallucinations: DETECTED: No
8. Information Processing Failures: DETECTED: No

COORDINATION ERRORS:
9. Task Derailment: DETECTED: No
10. Goal Deviation: DETECTED: No
11. Context Handling Failures: DETECTED: No
12. Verification Failures: DETECTED: No

TASK_STATUS: Needs correction
OVERALL_GUIDANCE: You are repeating actions unnecessarily and using incorrect tools. Specifically:
1. Stop running the same test command repeatedly - the test 'pytest test_file.py' has already been executed successfully three times with the same result
2. Use the Python interpreter tool for executing Python code, not the text editor which is only for viewing and modifying files
3. Now focus on analyzing the test results you already obtained to determine what the next development step should be
4. Review the test output to identify any failing tests or areas that need improvement
5. If all tests are passing, proceed to verify your implementation meets the original requirements before considering the task complete

=====================================================
INSTRUCTIONS
=====================================================

1. Focus on errors that can be corrected through guidance
2. Provide specific, actionable recovery instructions
3. Be concise but precise in evidence citations
4. Only mark "DETECTED: Yes" if you have clear evidence
5. Prioritize errors that most threaten task completion

Now review the agent's trajectory and provide your supervisor feedback.
\end{lstlisting}

\subsection{Budget-Aware Instructions}
\label{app:step_aware}

After the agent passes step 100 of its 150-step budget, the following budget-aware instructions are appended to whichever critic prompt is in use:

\begin{lstlisting}[basicstyle=\ttfamily\scriptsize,xleftmargin=-4pt,xrightmargin=-4pt]
NOTE: The agent has used {{current_step}} of its {{step_limit}} total steps. Work that is not submitted before the step limit will be permanently lost. If the agent has made code changes that address the issue, prioritize recommending submission. If the agent appears stuck in a loop, recommend that it submit its current changes immediately rather than continuing to iterate.
\end{lstlisting}

\section{Opus Teacher Prompt Ablation}
\label{app:opus_prompt_ablation}

Table~\ref{tab:opus_prompt_ablation} compares detailed and high-level feedback from the \opus{} teacher on the two Qwen agents.

\begin{table*}[t]
\centering
\small
\setlength{\tabcolsep}{5pt}
\resizebox{\textwidth}{!}{%
\begin{tabular}{@{}llrrrrrr@{}}
\toprule
\textbf{Agent} & \textbf{Opus Prompt} & \textbf{Resolved \%} & \textbf{Localized \%} & \textbf{Non-Empty \%} & \textbf{Non-Loop \%} & \textbf{Avg.\ Steps} & \textbf{Total Cost} \\
\midrule
\multirow{2}{*}{\qwenThirtyTwoB{}}
  & High-Level & 26.8 & 67.8 & 82.4 & 92.4 & 25.0 & \$0.267 \\
  & Detailed & \textbf{40.6} & \textbf{72.6} & \textbf{85.0} & \textbf{93.8} & 24.9 & \$0.337 \\
\midrule
\multirow{2}{*}{\qwenNext{}}
  & High-Level & 38.2 & 81.4 & 100.0 & \textbf{99.4} & 17.0 & \$0.172 \\
  & Detailed & \textbf{48.0} & \textbf{85.0} & 100.0 & 98.6 & 18.1 & \$0.241 \\
\bottomrule
\end{tabular}
}
\caption{\opus{} teacher prompt comparison on SWE-Bench Verified (500 instances). \emph{Non-Loop} is the percentage of trajectories without three consecutive repetitions of the same Bash command. Total cost is per-instance agent-plus-critic cost. Bold marks the better percentage metric within each agent; ties are unbolded.}
\label{tab:opus_prompt_ablation}
\end{table*}

\section{Inference Pricing}
\label{app:pricing}

All per-instance dollar costs reported in this paper are computed from input/output token counts using the public hosted-inference rates in Table~\ref{tab:pricing}. Qwen-family rates are taken from PricePerToken. \cwm{} is not publicly hosted, so we use \qwenThirtyTwoB{} as a proxy on the basis that both are dense 32B models. The same \qwenEightB{} rate is applied to the untrained \qwenEightB{} critic and to all \qwenEightB{} SFT variants distilled from it.

\begin{table*}[t]
\centering
\small
\setlength{\tabcolsep}{4pt}
\begin{tabular}{@{}l r r l@{}}
\toprule
\textbf{Model} & \textbf{Input (\$/Mtok)} & \textbf{Output (\$/Mtok)} & \textbf{Source} \\
\midrule
\cwm{}                              & 0.080 & 0.280 & Proxy: \qwenThirtyTwoB{} (same-size 32B dense) \\
\qwenThirtyTwoB{}                   & 0.080 & 0.280 & PricePerToken\footnote{\url{https://pricepertoken.com/pricing-page/model/qwen-qwen3-32b}} \\
\qwenNextEightyB{}                  & 0.090 & 0.780 & PricePerToken\footnote{\url{https://pricepertoken.com/pricing-page/model/qwen-qwen3-next-80b-a3b-instruct}} \\
\qwenEightB{} and SFT variants      & 0.050 & 0.200 & PricePerToken\footnote{\url{https://pricepertoken.com/pricing-page/model/qwen-qwen3-8b}} \\
\opus{}                             & 5.500 & 27.500 & API pricing \\
\bottomrule
\end{tabular}
\caption{Hosted-inference rates used to convert token counts to the dollar costs in this paper. Opus critic costs include prompt-caching: cache-write tokens billed at $1.25\times$ input (\$6.875/Mtok) and cache-read at $0.10\times$ input (\$0.55/Mtok); the reported Opus cost combines these.}
\label{tab:pricing}
\end{table*}

\section{Statistical Significance}
\label{app:significance}

All six headline gains are statistically significant under a two-sided exact McNemar test on paired outcomes from the same 500 instances, and their paired 95\% confidence intervals exclude zero (Table~\ref{tab:significance}; $p<.001$).

\begin{table*}
\centering
\small
\setlength{\tabcolsep}{6pt}
\resizebox{\textwidth}{!}{%
\begin{tabular}{@{}llrrrr@{}}
\toprule
\textbf{Agent Model} & \textbf{Critic} & \textbf{Base (\%)} & \textbf{+ Critic (\%)} & \textbf{$\Delta$ 95\% CI} & \textbf{McNemar $p$} \\
\midrule
\texttt{Qwen3-32B}
  & \texttt{Qwen3-4B} + SFT + DPO & 8.8 & \textbf{14.4 (+5.6)} & $[+2.8,+8.4]$ & $1.3{\times}10^{-4}$\textsuperscript{***} \\
\texttt{Qwen3-Next-80B-A3B}
  & \texttt{Qwen3-4B} + SFT + DPO & 20.0 & \textbf{26.2 (+6.2)} & $[+3.2,+9.2]$ & $6.5{\times}10^{-5}$\textsuperscript{***} \\
\texttt{GPT-OSS-20B}
  & \texttt{Qwen3-4B} + SFT + DPO & 3.0 & \textbf{14.8 (+11.8)} & $[+8.6,+15.0]$ & $1.0{\times}10^{-12}$\textsuperscript{***} \\
\texttt{GLM-4.7-Flash-30B-A3B}
  & \texttt{Qwen3-8B} + SFT & 21.6 & \textbf{37.6 (+16.0)} & $[+12.1,+19.9]$ & $8.6{\times}10^{-15}$\textsuperscript{***} \\
\texttt{GPT-OSS-120B} (medium reasoning)
  & \texttt{Qwen3-4B} + SFT + DPO & 20.4 & \textbf{34.8 (+14.4)} & $[+10.6,+18.2]$ & $1.7{\times}10^{-13}$\textsuperscript{***} \\
\texttt{o3-mini}
  & \texttt{Qwen3-8B} + SFT & 19.0 & \textbf{29.4 (+10.4)} & $[+7.0,+13.8]$ & $3.1{\times}10^{-9}$\textsuperscript{***} \\
\bottomrule
\end{tabular}%
}
\captionof{table}{Statistical significance of the headline resolve-rate gains in Table~\ref{tab:main} on SWE-Bench Verified. Parentheses show percentage-point gains over no critic; 95\% CIs are paired McNemar--Wald intervals on these gains. All critic runs use $k=5$. Significance markers: \textsuperscript{***}$p<.001$, \textsuperscript{**}$p<.01$, and \textsuperscript{*}$p<.05$.}
\label{tab:significance}
\end{table*}

\section{Qualitative Case Studies}
\label{app:qualitative}

Verbatim critic excerpts cited in §\ref{sec:qualitative}; ellipses mark omitted rubric boilerplate (\S\ref{app:critic_prompts}).

\subsection{Untrained vs.\ trained critic on the same loop}
\label{app:untrained_vs_trained}

On \texttt{django\_\_django-16595} the untrained critic flags only Verification Failures and prescribes a test the agent has not set up; the agent abandons self-correction and submits a corrupted file (Listing~\ref{lst:untrained_critic_excerpt}). On \texttt{django\_\_django-9296}, under matched surface conditions (\texttt{sed} edits to a Django source file), the trained critic flags Step Repetition and Tool Selection Errors jointly and prescribes a heredoc rewrite (Listing~\ref{lst:trained_critic_excerpt}); the agent submits a correct patch ten steps later.

\begin{lstlisting}[caption={Untrained critic at agent step 10 on \texttt{django\_\_django-16595}. Only Verification Failures is flagged, with a recovery prescribing a test the agent has not set up.}, label={lst:untrained_critic_excerpt}]
... 12. Verification Failures: DETECTED: Yes
EVIDENCE: "The agent made changes to the reduce method in
django/db/migrations/operations/fields.py without verifying
if the fix works"
RECOVERY_ACTION: "Run the test script to verify that the fix
resolves the issue with multiple AlterField operations not
being reduced"

TASK_STATUS: Needs correction
OVERALL_GUIDANCE: You have modified the reduce method in the
AlterField class ... Run the test script to ensure that the
optimizer now correctly reduces multiple AlterField operations
... If the tests pass, you can submit your changes.
\end{lstlisting}

\begin{lstlisting}[caption={Trained critic at agent step 10 on \texttt{django\_\_django-9296}. Flags Step Repetition and Tool Selection Errors jointly and tells the agent to switch off \texttt{sed} append.}, label={lst:trained_critic_excerpt}]
3. Step Repetition: DETECTED: Yes
EVIDENCE: Agent has repeatedly applied the same sed append
command, each time creating duplicate `__iter__` methods
inside the `Paginator` class.
RECOVERY_ACTION: Stop using sed append commands and instead
rewrite the file using a heredoc or a different approach ...
6. Tool Selection Errors: DETECTED: Yes
EVIDENCE: Agent is using `sed -i '/def page_range(self):/a'`
... but this creates multiple duplicate method definitions
each time it's run because the pattern matches multiple
times after the first insertion.
RECOVERY_ACTION: Use a different editing strategy: write the
entire file using a heredoc, or use a Python script ...

TASK_STATUS: Critical intervention required
OVERALL_GUIDANCE: The file is now corrupted with multiple
duplicate `__iter__` method definitions. You must view the
file ... and then use a Python script (not sed append) to
properly insert a single `__iter__` method ...
\end{lstlisting}

\subsection{Routing around a dependency failure}
\label{app:sandbox_adaptation}

On \texttt{sympy\_\_sympy-22714} the unguided agent loops on \texttt{ModuleNotFoundError: mpmath}; the trained critic at step $20$ identifies the import as the root cause and tells the agent to bypass it (Listing~\ref{lst:sympy_excerpt}), and the agent submits a correct patch at step $25$.

\begin{lstlisting}[caption={Trained critic at agent step 20 on \texttt{sympy\_\_sympy-22714}.}, label={lst:sympy_excerpt}]
TASK_STATUS: Needs correction
OVERALL_GUIDANCE: The mpmath import in sympy/__init__.py is
preventing any import of sympy. You need to bypass this
import to test, or focus on reading the source code directly
to understand the issue without running code. The key issue
is in `sympy/core/sympify.py` line 472 where `sympify` calls
`parse_expr` with `evaluate=evaluate` ...
\end{lstlisting}

\subsection{Backseat driving in the detailed teacher prompt}
\label{app:backseat_driving}

The detailed prompt at the \opus{} teacher mixes patch dictation and architectural prescription in the same firing slot. Listing~\ref{lst:django12193_excerpt} is the dictation case on \texttt{django\_\_django-12193} ($\rho{=}1.00$, resolved): \opus{} supplies the before/after blocks for \texttt{CheckboxInput.get\_context} and the agent transcribes them into a patch over its next three messages. Listing~\ref{lst:sphinx7590_excerpt} is the wrong-fork case on \texttt{sphinx-doc\_\_sphinx-7590} ($\rho{=}0.70$, gold overlap $0.12$, fails): \opus{} hedges between modifying the existing parser path and introducing a new AST class, the agent commits to the latter and writes $46$ lines around an \texttt{ASTUserDefinedLiteral} class that the gold patch never introduces.

\begin{lstlisting}[caption={\opus{} teacher (detailed prompt) at agent step $6$ on \texttt{django\_\_django-12193}. The agent's submitted patch reproduces both blocks verbatim.}, label={lst:django12193_excerpt}]
OVERALL_GUIDANCE: The agent has correctly identified the
root cause in `CheckboxInput.get_context()` ... The fix
should be to make a copy of `attrs` before modifying it.
Specifically, change the `get_context` method of
`CheckboxInput` from:

```python
def get_context(self, name, value, attrs):
    if self.check_test(value):
        if attrs is None:
            attrs = {}
        attrs['checked'] = True
    return super().get_context(name, value, attrs)
```

to:

```python
def get_context(self, name, value, attrs):
    if self.check_test(value):
        attrs = {**(attrs or {}), 'checked': True}
    return super().get_context(name, value, attrs)
```
\end{lstlisting}

\begin{lstlisting}[caption={\opus{} teacher (detailed prompt) at agent step $6$ on \texttt{sphinx-doc\_\_sphinx-7590}. Bullet $4$'s ``or if a new \dots\ class is needed'' is the wrong fork; the agent takes it.}, label={lst:sphinx7590_excerpt}]
3. Implementation approach: After matching a numeric
literal and consuming any standard suffixes (`uUlLfF`),
also try to match an identifier suffix (the ud-suffix)...

4. Check the AST classes - look at `ASTNumberLiteral`,
`ASTStringLiteral`, `ASTCharLiteral` to see if they need
modification, or if a new `ASTUserDefinedLiteral` class
wrapping them is needed.
\end{lstlisting}

% \subsection{Where the trained critic still falls short}
% \label{app:fall_short}

% Although the trained critic improves resolve rate across agent models, it can still induce repeated actions on some trajectories. In these cases, the critic identifies the loop but reissues near-identical \texttt{OVERALL\_GUIDANCE} every $k$ steps, which the agent stops acting on until the budget is exhausted (App.~\ref{app:repeated_demand}). The SFT loss matches teacher tokens at each prefix without penalizing the agent-side outcome, which we discuss further in §\ref{sec:discussion}.

\subsection{Repeated demand on \texttt{django\_\_django-13012}}
\label{app:repeated_demand}

Listing~\ref{lst:repeated_demand} shows two of the $13$ firings on \texttt{django\_\_django-13012} to illustrate the structural similarity of the \texttt{OVERALL\_GUIDANCE} that the agent stops acting on.

\begin{lstlisting}[caption={Trained critic firings $6$ (msg 127) and $10$ (msg 211) on \texttt{django\_\_django-13012}; both prescribe the same fix path.}, label={lst:repeated_demand}]
[firing 6, agent step 60]
TASK_STATUS: Critical intervention required
OVERALL_GUIDANCE: You are stuck in a loop viewing the same
code. The `ExpressionWrapper.get_group_by_cols` calls
`self.get_source_expressions()` which returns
`[self.expression]`, and then calls `source.get_group_by_cols()`
on that expression. When `self.expression` is a `Value`,
`Value.get_group_by_cols` returns `[]`, so the constant
should NOT be in GROUP BY. The real issue is likely in
`query.py`'s `set_group_by` method ...

[firing 10, agent step 100]
TASK_STATUS: Critical intervention required
OVERALL_GUIDANCE: You are stuck in a loop viewing the same
code. Create a minimal reproduction script that only imports
`Value`, `IntegerField`, and `Sum` from `django.db.models`
(no `functions` or `asgiref`), then look at `set_group_by`
in `query.py` (line 1942) to understand how the GROUP BY
clause is populated ...
\end{lstlisting}

\end{document}